\documentstyle[eqsecnum,aps,multicol,epsfig,prb]{revtex}

\begin{document}

\title{Transport and magnetic anisotropy  in CMR thin film \\
La$_{1-x}$Ca$_{x}$MnO$_{3}$ ($x \simeq 1/3$)\\
induced by a film-substrate interaction}

\author{B. I.  Belevtsev$^{*}$ and V. B. Krasovitsky}
\address{B. Verkin Institute for Low Temperature 
Physics \& Engineering, Kharkov, 61164, Ukraine
}
\author{D. G. Naugle$^{\dag}$, K. D. D. Rathnayaka and A. Parasiris}
\address{Department of Physics, Texas A\&M University, 
College Station, TX 77843-4242, USA
}
\author{S.R. Surthi and R.K. Pandey}
\address{The Department of Electrical and Computer Engineering,
The University of Alabama, Tuscaloosa, AL 35486, USA}

\author{M. A. Rom}
\address{Institute for Single Crystals, Kharkov, 61001, Ukraine}
\maketitle
\begin{abstract}
We present a study of anisotropy of transport and magnetic properties
in a La$_{1-x}$Ca$_{x}$MnO$_{3}$ ($x \approx 1/3$) film prepared by 
pulsed-laser deposition onto a LaAlO$_{3}$ substrate. We found a
non-monotonic dependence of magnetoresistance (MR) on magnetic field $H$ for 
both $H$ perpendicular and parallel  to the film plane but  
perpendicular to the current.  In the longitudinal geometry (when $H$ is 
parallel to both the current and the film plane) the MR was negative at all
fields below 20 kOe, 
as expected for colossal-magnetoresistance manganites. This rather complex 
behavior of MR manifests itself at rather low temperatures, far below the 
Curie temperature $T_c$, which was close to room temperature. Two main sources 
of MR anisotropy in the film have been considered in the explanation of the 
results: (1) the existence of preferential directions of magnetization (due 
to strains stemming from the lattice film-substrate mismatch or other 
reasons); (2) dependence of resistance on the angle between current and 
the magnetization, which is inherent in ferromagnets. The transport and 
magnetic properties of the film correspond well to this view. In particular, 
the following angle dependence of MR is found:
$R({\theta})/R(0) = 1 + \delta_{an}(T,H)\sin^{2}\theta$ (where $\theta$ is 
the angle between the field and current directions in the plane normal to the
film but parallel to the current). The temperature
and magnetic field dependences of $\delta_{an}(T,H)$ were recorded and
analyzed. A clear magnetization anisotropy, that generally 
favors the magnetization in the film plane is also found. At the same time the 
recorded
magnetization curves (as well as the MR data) indicate, that the film 
crystal structure should be inhomogeneous in such a way that various parts 
of the films have non-identical magnetic properties (with different
directions of spontaneous magnetization). This hypothesis is supported by 
X-ray diffraction which revealed that the film is inhomogeneous in strain,
lattice parameter and lattice orientation. This peculiar macroscopic-scale 
disorder is caused by a film-substrate interaction.  The possible reasons for 
formation of such structure and its effect on MR anisotropy are considered.
\end{abstract}
\begin{multicols}{2}
\section{Introduction}
Discovery of so called colossal magnetoresistance (CMR) in doped lanthanum 
manganites of the type La$_{1-x}$A$_{x}$MnO$_{3}$ (where $A$ is a divalent 
alkaline-earth element like Ca, Sr, Ba) \cite{helm93,jin94} has fundamental 
importance for solid state physics and offers promissing application in 
advanced technology. The most pronounced effect of CMR ({\it i.e.}, an 
extremely large negative magnetoresistance (MR)) was found in 
La$_{1-x}$Ca$_{x}$MnO$_{3}$ films with $x \simeq 1/3$. In the concentration 
range $0.2 < x < 0.5$ the La$_{1-x}$Ca$_{x}$MnO$_{3}$ is a ferromagnet with 
a rather high conductivity at temperatures far below the Curie temperature
$T_c$. The resistance strongly increases with temperature and has a peak at  
a temperature $T_{p}$ which is close to $T_{c}$ in samples with fairly 
perfect crystalline structure.  The MR is maximal near $T_c$. 
Above $T_{p}$ (in the paramagnetic state) the resistance has semiconducting 
behavior and the MR is much less. Below $T_{c}$ the MR strongly decreases 
with decreasing temperature, and it is believed that the MR must go to zero 
on approaching $T = 0$~K in fairly good crystals\cite{hund,gupta96}.    
\par
The common explanation of the CMR is usually provided in the frame 
of the double-exchange (DE) model \cite{zener,and55,gennes} which is 
based on the assumption of the appearance of Mn$^{4+}$ ions with substitution 
of La$^{3+}$ by a divalent cation. It is believed that in this case 
a ferromagnetism results from the strong ferromagnetic exchange
between  Mn$^{3+}$ and Mn$^{4+}$. This model, however, cannot
explain the many features of the resistivity behavior of manganites in
both the ferromagnetic and the paramagnetic states. Therefore, in succeeding 
theoretical 
works additional physical mechanisms (mainly still in the frame of DE 
model) were considered. The possible influence of strong 
electron-phonon coupling (Jahn-Teller distortion), polaronic effects 
(magnetic or lattice polarons), nearly half-metallic nature of 
ferromagnetism in the manganites, electron localization, phase separation and 
other effects  were considered (see Refs.\ 
\onlinecite{nagaev,ramirez,khom,coey2,dagot} and references therein). 
\par
In spite of extensive experimental and theoretical efforts a clear 
understanding of CMR in the manganites is not yet available. The reason is 
that the 
knowledge of even the basic electronic properties of doped manganites is still 
far from complete.  For example, one can find in
the literature conflicting experimental claims regarding the nature of 
holes in doped manganites at $x < 0.5$.  
The DE model is based on the assumption that the holes in doped manganites 
correspond to Mn$^{4+}$ ions arising among the regular Mn$^{3+}$ ions 
due to the doping. But some investigations give strong evidence that the 
holes are located mainly on the oxygen ions \cite{saitoh,ju} ({\it i.e.}, the 
holes are of O $p$ character).  On the other hand there is experimental 
evidence (see Ref.\ \onlinecite{croft} and references therein) that holes 
doped into LaMnO$_{3}$ are mainly of Mn $d$ character. 
\par
One of the important questions in physics of the CMR manganites is the 
nature of the rather high-conducting state below $T_{c}$. The only 
sure assumption at present is that the charge carriers at low temperatures 
can be considered to be quasifree. Whether the doped CMR manganites in the
ferromagnetic state should be regarded as conventional bad (disordered) metals 
or as just heavily doped degenerate semiconductors has been argued 
\cite{nagaev,pickett,mahen}. It is known, however, 
that manganites do not behave like conventional non-ferromagnetic metals. For 
example, the  decrease in resistivity with decreasing temperature 
 of 
fairly good crystalline manganites is too large to 
be attributed, as in conventional metals, to the electron-phonon interaction 
\cite{mahen}. It follows also from the known experimental 
data\cite{hund,font} that a  clear correlation exists between transport 
properties and magnetism in doped manganites. Namely, the resistance $R$ of 
manganites in the ferromagnetic state is a function of the magnetization $M$ 
which 
in turn depends on the temperature and magnetic field: $R = f[M(T,H)]$. 
In manganites the conductivity increases with the enhancement of 
ferromagnetic order. This is actually the source of the huge resistivity
decrease at the paramagnetic-ferromagnetic transition and the CMR.
This correlation is most pronounced in good quality crystals, but persists 
to some degree even in rather disordered systems, like polycrystalline or 
granular samples. 
\par
The bulk manganites La$_{1-x}$A$_{x}$MnO$_{3}$ ($x \simeq 1/3$) have nearly 
cubic symmetry and therefore should not have any marked MR anisotropy. 
In contrast, the CMR films possess a pronounced MR anisotropy in low 
magnetic fields \cite{eck,li,wangli}. Due to the above-mentioned 
transport-magnetism correlation it should be thought that the MR anisotropy in 
CMR films is in fact some reflection of $M(T,H)$ behavior. Two main sources 
of MR anisotropy in ferromagnetic films are: (1) the existence 
of preferential directions of magnetization (due to strains stemming from the 
lattice film-substrate mismatch or other sources), and 
(2) dependence of resistance on the angle between current and 
magnetization, which is inherent in ferromagnets (the so called anisotropic 
magnetoresistance (AMR) effect)\cite{Mc,dahl}.
\par        
It was found\cite{eck,wangli} for the CMR films, which are subject to 
compressive strain in the film plane, that if the easy magnetization axis is 
parallel to the film plane, an unusual  positive MR appears when the 
magnetic field is perpendicular to the film 
plane, while for a parallel field the MR is negative. 
This behavior can be associated with concurrent influence of the 
above-mentioned anisotropy sources\cite{eck,wangli}. In this article, 
we report that MR anisotropy in CMR films can also manifest itself in far more 
complex and puzzling ways. The object of study was La$_{1-x}$Ca$_{x}$MnO$_{3}$ 
($x \approx 1/3$) films prepared by pulsed-laser deposition (PLD) onto 
LaAlO$_{3}$ substrates. We found non-monotonic and alternating dependences 
of the MR on $H$ for both the perpendicular and parallel directions of $H$ 
relative to the film plane with $H$ perpendicular 
to the current (preliminary short report about this behavior was presented 
at LT22\cite{bel1}). Only in the longitudinal geometry (when 
$H$ is parallel to both  the current and the film plane) was the MR always 
negative, as expected for CMR manganites. This rather complex behavior of MR 
manifests itself in the ferromagnetic state and has not been reported in 
previous studies.  
We will show below that this behavior is determined by a peculiar structural
disorder induced by a film-substrate interaction. From the transport and 
magnetic properties of the film studied, it can be concluded that the film 
crystal structure should be inhomogeneous in such a way that various 
parts of the film have non-identical (and quite distinct) magnetic 
properties. This hypothesis is supported by x-ray 
diffraction which revealed that the film is inhomogeneous in strains,
lattice parameters and lattice orientation. The possible reasons for the
formation of such structure and its effect on MR anisotropy are considered.
We note that although similar MR behavior was observed for several films
prepared by this technique, the detailed measurements reported here were all
taken on the same film, a representative specimen.

\section{Experiment}
La$_{1-x}$Ca$_{x}$MnO$_{3}$ ($x \approx 1/3$) films 
were grown  by PLD on (100) oriented LaAlO$_3$ 
substrate. A PLD system from Neocera Inc. with a Lambda Physik KrF excimer 
laser operating at 248~nm was used to ablate the target material with
a nominal composition La$_{2/3}$Ca$_{1/3}$MnO$_{3}$. 
The main details of the target preparation and laser ablation technique are 
described in Ref.\ \onlinecite{pandey1}. 
Stoichiometric amounts of high purity La$_2$O$_3$, CaO and MnCO$_3$  
were mixed and ball milled for several hours, reacted at 1100$^{\circ}$C 
for 24 hours with intermediate grinding and mixing after 12 hours, and 
pressed with Duramax B-1020 acrylic binder at 50-170 MPa to make the target 
pellet. The pellet was sintered at 600 -- 1200$^{\circ}$C for 12 hours  
in a box furnace in air to burn off the binder and strengthen the pellet. 
During deposition the pulse energy  was 228 mJ with a repetition rate of 8 Hz. 
The target-substrate distance was about 7 cm. 
\par
The film 80 nm thick\cite{bel2} described in this paper was ablated at a
substrate temperature of 400$^{\circ}$C in an oxygen atmosphere at pressure 
$P_{O2} = 200$~mTorr. 
Time of deposition was about 20~min. Immediately after deposition the film 
was annealed 30 min at $T=400^{\circ}$C in the same PLD chamber at  
$P_{O2} = 330$~mTorr. The film was also post-annealed in flowing oxygen 
for 24 hours at 900$^{\circ}$C. 
\par
X-ray diffraction (XRD) study of crystal-structure  of the film 
and the substrate was done using a DRON-3 diffractometer with a Ge(111) 
monochromator and CuK$_{\alpha 1}$ radiation. Magnetization and ac 
susceptibility measurements were done by commercial SQUID magnetometer. 
Resistance as a function of field and temperature  was measured using a 
standard four-point probe technique. The available cryostat with a rotating 
electromagnet makes it possible 
to measure resistance in magnetic fields up to 20 kOe with different directions of
$H$ relative to the plane of the film and the transport current. 

\section{Results}
\subsection{X-ray diffraction}
\label{xray}
\subsubsection{Analysis method}
The two methods of XRD study were used: (i) normal $\Theta - 2\Theta$ 
scanning, and (ii) diffractional reflection curve (DRC) recording. 
DRCs were recorded on symmetric and asymmetric 
reflections. The technique of sample rotation about the diffraction vector 
was used\cite{rom}. That makes variations of the angle between the 
surface and incident beam or corresponding reflected one possible up to the 
critical angle of total external reflection which is about $\simeq 1^\circ$. 
The perfection of crystal structure was characterized by the DRC half-width 
$\beta_{1/2}$. The crystal lattice parameters were obtained by the Bond 
 technique\cite{bond}. 

\subsubsection{Substrate characterization}
The substrate LaAlO$_3$ (from Coating \& Crystal Technology (CCT), Kittaning,
PA 16201) was (001) oriented. For determination of lattice parameters 
$a_s, b_s, c_s$ (of the pseudocubic cell) the reflections (400), (330) and (003) 
were used. It was found that $a_s = 0.37902$~nm,  $b_s = 0.37956$~nm, and 
$c_s = 0.37836$~nm. One can compare these values with CCT data 
($a = 0.379$~nm) or with that from one of the special studies 
of LaAlO$_3$ single crystals\cite{surf} ($a = 0.37896$~nm). The substrate is 
characterized by the availability of mosaic crystal blocks and twin structure 
which are common to LaAlO$_3$\cite{surf}. The angles of misalignment of 
fragments (estimated with use of the asymmetric reflections)  range up to 
0.2$^{\circ}$. The magnitudes of 
$\beta_{1/2}$ for these fragments are dispersed between 15 and 120
arcsec for different parts of the crystal. Using the asymmetric reflection 
(101) enables us to conclude that DRC does not experience any significant 
broadening even at minimum angles ($\approx 1^\circ$) of reflected beam. 
This demonstrates that the PLD process in our case does 
not involve a formation of a damage layer on the substrate surface in 
contrast to the study of Ref.\ \onlinecite{rom} where 
the value of $\beta_{1/2}$ for the NdGaO$_3$ substrate was increased by two 
orders of magnitude, and the damage layer was 1.2~$\mu$m thick.   

\subsubsection{Film}
The film lattice parameters have been determined from the lines (040), 
(220) and (022). The in-plane lattice parameters are  found to be 
$a_f = 0.3836 \pm 0.0002$~nm and $c_f = 0.3831 \pm 0.0002$~nm; whereas, the 
out-of-plane one $b_f = 0.3867 \pm 0.0002$~nm. Hence the film has a 
tetragonal lattice.  The ratio of the out-of-plane
lattice parameter to the in-plane ones is about 1.009. Bulk 
La$_{1-x}$Ca$_{x}$MnO$_{3}$ ($x \approx 1/3$) has a cubic lattice with 
lattice parameter $a_p$ in the range 0.385--0.386~nm\cite{mahen,bulk}. 
It  is thus apparent that the film is in a strained state. In the plane of 
growth the film is under biaxial compression, but it is under uniaxial 
tension in the direction perpendicular to the film plane. Such a  strain state 
was observed previously in La$_{1-x}$Ca$_{x}$MnO$_{3}$ 
($x \approx 0.2$--0.3) films on  LaAlO$_3$ substrates\cite{koo,nath}. 
The effects connected with it will be discussed below. 
\par
It follows from the XRD data that the crystallographic substrate
plane (100)$_s$ (which corresponds to the deposition surface) is parallel
to the plane (010)$_f$ of the film. The fine-structure parameters were 
obtained from the analysis of several reflections taking into account the 
DRC broadening. The $\beta_{1/2}$ values for the reflections of (101), (202) 
and (303) are 0.375$^{\circ}$, 0.42$^{\circ}$, 0.56$^{\circ}$, 
correspondingly. The values of microblock angular misalignment, $\delta$, 
dimensions of coherent scattering areas, $L$, and microdeformation, 
$\varepsilon$, are  0.32$^{\circ}$, 60 nm, and $9\times 10^{-4}$,
correspondingly. 
\par
The DRCs at angles in the range 1--6$^{\circ}$ were asymmetric. There are 
reasons to believe that this is caused by superposition of reflections from 
parts  of the film with different lattice parameters, since  orientations
$[100]_{f}\parallel [010]_{s}$ and $[100]_{f}\parallel [001]_{s}$ are quite
possible due to the  twin structure of substrate. 
These XRD data will be used below for explanation
of the transport and magnetic anisotropy in the film.

\subsection{Transport and magnetic properties}

\subsubsection{General properties}
The transport properties of the film in Fig.1 correspond well to the 
expected behavior of CMR films \cite{helm93,jin94}. 
Namely, the temperature dependence of the resistance $R(T)$ has a maximum 
(peak) at $T_{p} \approx 296$~K. Below $T_{p}$ a quite sharp resistance 
drop takes place which corresponds to paramagnetic-ferromagnetic transition
that occurs approximately simultaneously with the insulator-metal transition. 
The resistivity $\rho$ at $T=4.2$~K was about 375 $\mu \Omega$cm. The ratio 
of the resistances at $T_p$ and 4.2 K, $R_{295}/R_{4.2}$, is about 31.4. 
This fairly large  variation of resistance with temperature for a rather 
disordered doped manganite should be attributed mainly to the strengthening 
of the magnetic order with decreasing temperature.  The magnetic field $H$ 
produces a large decrease in resistance (see the insert in Fig.1). For a 
measure of the MR we have taken $\delta_{H} = [R(0) - R(H)]/R(0)$. 
It was found that $\delta_{H}$ has its maximum absolute value (about 0.43 at 
 $H=16$~kOe) at a temperature $T_{m} \approx 272$~K. 
\par
The temperature dependences of the magnetization $M$ and of the AC 
susceptibility 
$\chi^{'}$ for different directions of magnetic field relative to the film 
plane are shown in Figs.~2 and 3. These enable the value of $T_c$ to be 
estimated.  Since the magnetic moment is the order 
parameter at a paramagnetic-ferromagnetic transition, it is quite 
natural to define $T_c$ as the temperature where $M$ or $\chi^{'}$ starts to 
increase, when going from high to lower temperatures. In this case (as may 
be seen in Figs.~2 and 3) the $T_c$ value of the film is 
approximately equal to the value of $T_p \approx 296$~K. 
\par
The nearly identical values of $T_c$ and $T_p$ are characteristic of 
films with good enough crystal perfection and fairly large grain 
size\cite{ara}.  If $T_c$ is defined as 
the temperature at which $M$ comes to a half of the saturation value, or 
as the temperature of the inflection point in $M(T)$, as is 
sometimes done, the value will 
appear to be somewhat lower (270--280 K). In any case, however, 
the value of $T_c$ and, especially, $T_p$ (which is determined quite 
unambiguously) seem to be somewhat higher than the corresponding values 
(260-270~K) for bulk Ca-doped manganites of the same composition 
($x = 1/3$) based on the 
accepted bulk phase diagram\cite{ramirez,coey2,schif}. An increase in $T_c$  
and $T_p$ was found earlier in bulk manganites under hydrostatic 
pressure\cite{ramirez,coey2,neu} or in films with considerable compressive 
strains due to a film-substrate interaction\cite{koo,shre}.  The maximal 
effect of hydrostatic pressure on bulk 
La$_{1-x}$Ca$_{x}$MnO$_{3}$ ($x \approx 1/3$) causes the increase of $T_p$ 
from $\approx 270$~K to 290 K, and $T_c$ from $\approx 268$~K to 
$\approx 285$~K\cite{neu}.  
\par
The total bulk strain in this film is highly compressive. Indeed, in 
this film the volume of the unit cell, $v_p$, is equal to 
$\approx 5.682\times 10^{-2}$~nm$^{3}$ which is less than $v_p$ for 
bulk La$_{2/3}$Ca$_{1/3}$MnO$_{3}$ (about $5.75\times 10^{-2}$~nm$^{3}$ 
provided $a_p = 0.386$~nm). Moreover, this is also less than previously reported 
values of $v_p$ for La$_{1-x}$Ca$_{x}$MnO$_{3}$ films\cite{koo,nath,shre}.  
It was found in preceding studies\cite{koo,shre} that $T_p$ (and $T_c$) 
increase with decreasing $v_p$ in La$_{1-x}$Ca$_{x}$MnO$_{3}$ films. 
For example, it was reported\cite{shre} that for a film with x=0.3 
and $v_p = 5.725\times 10^{-2}$~nm$^{3}$, $T_p$ is about 275 K . 
For the film of Figs.1--3 the unit cell volume is less, explaining  its higher 
$T_c$ (296~K). 
\par
The strain state of the film is however inhomogeneous. In this
case the influence of the Jahn-Teller part of the strain tensor on $T_c$ 
can be taken into account\cite{millis}, in principle. The equilibrium lattice 
parameter in bulk La$_{2/3}$Ca$_{1/3}$MnO$_{3}$ is not known, however, with 
necessary accuracy for quantitative consideration of this effect.

\subsubsection{Magnetoresistance anisotropy and AMR effect}
According to early studies \cite{helm93,jin94,hund,gupta96} the MR value 
in manganites decreases profoundly (but {\it remains negative}) below $T_{c}$ 
as the temperature is reduced. More recently\cite{eck,wangli}, it was shown
that MR in strained manganite films can be positive at low temperature.
We present below a far more complex behavior of $R(T,H)$ for this  
film (Figs. 4--6) which is determined by magnetic anisotropy induced 
by a film-substrate interaction. In these Figures, as well as in the 
following text of the paper, we will use designations $H_{\bot}$ and 
$H_{\|}$ for the cases of field $H$ applied perpendicular and parallel to 
the film plane, correspondingly.
\par
Let us consider, first of all, the MR behavior for the cases that the magnetic 
field is perpendicular to the current. Figs. 4 and 5 present the case for 
different field orientation ($H_{\bot}$ and $H_{\|}$). We will look more 
closely at the MR behavior at helium temperature. It can be seen, that, for 
increasing $H_{\bot}$, the MR is first negative (at $H_{\bot} \leq 4$~kOe), 
then positive (4 kOe $\leq H_{\bot} \leq 12$~kOe), and then negative again 
(Fig.4). For increasing $H_{\|}$, the MR is positive below 
$H_{\|} \simeq 6$~kOe and negative above it (Fig.5). Anisotropic 
behavior of this kind occurs only at low temperatures. At $T > 20$~K the MR 
is negative for both directions of the magnetic field. 
\par
If $H$ is parallel to both the current $J$ and the film plane, the MR 
is always negative (Fig. 6). It can be seen from Fig. 6 that MR values differ 
signicantly for the cases when $H$ is perpendicular and those  parallel to the 
transport current. This is because of a pronounced dependence of the resistance 
on angle $\theta$ between $H$ and the current $J$. 
One of the measured angular dependences is shown in Fig.~7. It is found
that $R(\theta)$ at high enough 
field (more than 8~kOe) can be described fairly well by the relation
\begin{equation}
R({\theta})/R(0) = 1 + \delta_{an}(T,H)\sin^{2}\theta, \eqnum{1}
\end{equation}
where $\delta_{an} = \{[R(90^{\circ})/R(0)] -1\}$ is some positive parameter. 
This is a manifestation of the AMR effect, which should be 
inherent in ferromagnets\cite{Mc,dahl}. In manganite films this effect was
previously reported in Refs.\onlinecite{eck,ziese,amaral}.   
\par 
The magnetic field dependence of the AMR parameter $\delta_{an}$ at $T=4.2$~K
is shown in Fig. 8. It can be seen that $\delta_{an}$ 
increases with field in the range below $\approx 8$~kOe and comes to some 
saturation value at higher field. This saturation, as will be shown below,
proceeds approximately at the field $H_{s}$ where the magnetization comes to 
rotational saturation in a field perpendicular to the film plane. Some 
clear inflection in the $\delta_{an}(H)$ dependence at $H\approx 2.5$~kOe 
can be seen. This inflection should not take place for homogeneous systems 
(compare, for example, with results of Ref.\ \onlinecite{eck}) and reflects,
as will be shown below, the structural and magnetic inhomogeneity of the film
studied. 
\par
A temperature dependence of saturated values of $\delta_{an}$ is shown 
in Fig. 9. Three important features of $\delta_{an}(T)$ behavior can be 
derived from the Figure. First, $\delta_{an}$ is nearly constant at low 
temperatures (up to 150 K), where the saturation magnetization of the film 
does not depend practically on the temperature, as will be shown below.
Second,  $\delta_{an}$ values go up in the temperature range of 
the ferromagnetic-paramagnetic transition in such way that the $\delta_{an}(T)$ 
curve has a maximum at $T = 280$~K which is close to $T_c$ and $T_p$ 
(see Figs. 1--3). Third, the magnitude of $\delta_{an}$ goes clearly to zero at
the transition to the paramagnetic state. The latter is quite expected since the 
AMR effect is unique to the ferromagnetic state. 
\par
Let us present now a behavior of the ratio of resistances in magnetic 
fields perpendicular and parallel to the film plane (we denote these 
resistances as $R_{\bot}$ and $R_{\|}$) in the case that 
both fields are perpendicular to the current (Fig. 10). In this case the AMR 
effect has no influence on MR.  It can be seen that 
the ratio of $R_{\bot}$ and $R_{\|}$  is less than unity in low field 
($\lesssim 5$~kOe) and more than unity at higher field with a tendency
to saturation at high enough fields. This behavior should reflect the 
magnetization anisotropy behavior. In Fig.~11 the temperature dependence of 
the ratio between MR in parallel and that in perpendicular fields, 
$\Delta R_{\|}(H)/ \Delta R_{\bot}(H)$, is presented at $H = 15$~kOe 
(this field is high enough to saturate the magnetization in any direction, 
therefore, both the $\Delta R_{\|}(H)$ and $\Delta R_{\bot}(H)$ are negative). 
It is seen that MR anisotropy is high below $T \approx 30$~K. Then, with  
increasing temperature, $\Delta R_{\|}(H)/ \Delta R_{\bot}(H)$ decreases to 
some nearly constant value about 1.3, which persists up to 220 K. It decreases  
further at higher temperatures, reaching $\simeq 1$ at
room temperature, {\it i.e.} near the Curie temperature $T_c$. Therefore, this 
type of MR anisotropy is connected with ferromagnetic state as well. This 
point will be discussed in more detail below. 

\subsection{Magnetic anisotropy}
\label{magani}
The behavior of these MR anisotropy effects suggests that the anisotropy is 
caused not only by the AMR effect, but that it also results from the 
magnetization anisotropy. This anisotropy exists in this film as  shown 
in Figs. 2 and 3.  It is obvious from these Figures that the film is magnetized 
more easily in the field direction parallel to the film plane. To check more 
thoroughly, we 
have measured $M(H)$ in the both directions at different 
temperatures in the range 4--100 K in fields up to 20 kOe. We have 
found that the saturation magnetization $M_s$ essentially does not depend on 
the temperature in this range ($M_s$ is lowered only by a few percent after 
warming from 4 K to 100 K). The $M(H)$ dependences are observed to be quite
different for the field directions parallel and perpendicular to the film 
plane, but for the same field direction the recorded $M(H)$ curves 
for different temperatures practically merge together. Only in the low fields 
($\lesssim 0.5$~kOe in the parallel direction and $\lesssim 2.5$~kOe in the 
perpendicular direction) can differences be found 
between the low and high temperature behavior. To consider this and other 
effects more properly, Fig.~12 shows $M(H)$ dependences 
for $T=4$~K and 100~K. These two graphs illustrate low temperature
(below 30 K) and high temperature (above 50 K) peculiarities of $M(H)$ 
behavior for this same film.  First of all, it should be noted, 
that experimental points in these graphs present the data recorded for  
increasing and subsequently for decreasing applied magnetic field. No 
significant 
hysteresis in the $M(H)$ curves at low temperature and only a rather weak one 
at high temperature in weak fields can be seen. Second, there is a pronounced 
difference in the $M(H)$ curves recorded for parallel and perpendicular 
directions of the magnetic field, which provides an unquestionable evidence of 
the magnetization anisotropy. 
The $M(H)$ curves for both field directions are rather extended (therefore, 
it can  be argued that neither  represents the easy magnetization axis),
but it is, however, clear that, on the whole,  the film is magnetized more easily in the 
parallel field direction. 
\par
One more difference can be seen from comparing the two graphs in Fig. 12. 
Namely,
at low temperatures it appears as if the magnetization is nonzero and rather 
high at zero field for both field directions [Fig. 12 (a)]. For increasing
field, the magnetization increases, however, rather slowly.  By contrast, at 
high temperatures the magnetization increases with the field more
gradually, beginning from zero, without any peculiarities in low fields 
[Fig.~12~(b)].
\par 
It should be recognized, however, that true zero magnetic field cannot be set 
in the magnetometer. For experimental reasons there is always some stray 
magnetic field of the order of 1 Oe. This weak field is sometimes quite
enough to cause a significant magnetization at nominal $H = 0$ in the
case of low coercivity. The $M(H)$ behavior shown in Fig. 12 (a) suggests 
that, at low temperatures for both perpendicular and parallel field
directions, some parts of the film have a very low coercivity (this causes the
jump-like increase in the magnetization in very low fields); whereas, other parts
of the film have a higher coercivity. In other words, it can be argued that 
some parts of the film have a substantial in-plane magnetization; whereas, 
other parts have a substantial out-of-plane magnetization for a magnetic field 
which is very close to zero. This is an evidence that the demagnetization energy at low 
temperatures cannot overcome entirely the spontaneous (parallel and 
perpendicular) magnetization in some (different) parts of the film. This effect
is not pronounced at higher temperatures ($T>50$~K) [Fig.~12~(b)].
\par
It should be noted, that some features of the magnetic-anisotropy behavior of 
this film correlate well with its MR behavior. It follows from 
Fig. 11, that the absolute values of negative MR in parallel field are 
greater than those of in the perpendicular 
field ($\Delta R_{\|}(H)/ \Delta R_{\bot}(H) > 1$). Since the conductivity of
manganites increases with an enhancement of the magnetic order, this behavior
just reflects the point that the magnetization increases more easily in a 
magnetic field parallel to the film plane. This MR anisotropy is connected with
the ferromagnetic state. For this reason it disappears when $T$ approaches
$T_c$ (Fig. 11).

\section{Discussion}
In this Section we will discuss the different sources of MR 
anisotropy in ferromagnets (FM) and their possible effects in this film.   We 
will not consider here the influence of ballistic mechanisms of 
the MR and MR anisotropy in FM \cite{vons}, which are connected with the 
curving of electron trajectories in a magnetic field. These are important only if 
the electron mean-free path is fairly large, which is not the case in rather 
resistive manganites.  The AMR effect which is an intrinsic source of MR
anisotropy in any FM will be considered as the first point. This effect plays 
a crucial part in the MR anisotropy of the films studied. As the last but not 
least point, the extrinsic or induced sources of MR anisotropy which are 
caused by the shape and strain state of the FM will 
be thoroughly discussed. These are especially important for films and small 
particles. It will be shown that the rather complicated MR anisotropy
behavior found in this film can be explained by the concurrent influence of 
these intrinsic and extrinsic sources of MR anisotropy.

\subsection{AMR effect}
This effect in ferromagnets is thought to be caused by the spin-orbit 
interaction (see Refs.\ \onlinecite{Mc,dahl} and  references therein). The 
known theoretical models are related to 3${\it d}$ metals such as Ni, Co, Fe 
and its alloys. Some attempts to apply the similar model concepts to 
manganites were made in Ref.\ \onlinecite{ziese}. It can be said at the moment, 
however, that mechanisms for AMR in manganites are not clearly understood. From 
the other side, 
the essential features of this effect in manganites are already established 
rather well. The temperature dependence of the AMR parameter $\delta_{an}(T)$ 
recorded in this study (Fig. 9) corresponds well to previous 
results\cite{ziese,amaral}. Among other factors, the most important features of 
$\delta_{an}(T)$ behavior such as the constancy the magnitude of 
$\delta_{an}(T)$  
at low temperatures ($T \ll T_c$),  the maximum in $\delta_{an}(T)$ at 
$T \simeq T_c$, and the  approach to zero of $\delta_{an}(T)$ at $T>T_c$, 
correspond to the above-mentioned results.
\par
There is no clear notion among scientists at present as to which factors 
determine the magnitude of the AMR effect in manganites. Most of the authors 
usually refer to  behavior of 3$\it d$ metals and the corresponding theories 
developed for these metals. But this approach does 
not appear to be very fruitful. Indeed, it follows from the 3$\it d$-metal 
models\cite{Mc,dahl} that the AMR parameter $\delta_{an}$ should depend somehow 
on the magnetization.  Really, the AMR effect takes place only in 
ferromagnets with a spontaneous magnetization. After the transition to the 
paragmanetic state, and the  disappearence of the spontaneous magnetization,  
$\delta_{an}$ goes to zero. But no clear correlation between the AMR amplitude 
and the saturation magnetization was found for 3$\it d$ metals\cite{Mc}. Thus 
this type of general explanation about the  influence of magnetization is not 
very productive for understanding
the AMR effect in manganites. Let us consider an other example. It was
experimentally found for 3$\it d$ metals (with corresponding theoretical 
support)\cite{stampe} that $\delta_{an}$ is proportional to  
$T_c-T$ on approaching the Curie temperature from below. This means a linear 
decrease in $\delta_{an}$  to zero when $T$ approaches $T_c$. But this is 
clearly not the case for manganites where 
$\delta_{an}(T)$ has a maximal amplitude near $T_c$ (see Fig. 9 of this paper 
and corresponding Figures in Refs.\ \onlinecite{ziese,amaral}). Therefore, the 
AMR behaviors of manganites and 3$\it d$ metals are drastically different and
must be governed by different mechanisms. 
\par
Upon a closer view of results of this study together with these of Ref.\ 
\onlinecite{ziese,amaral} a clear correlation between the magnitudes of 
$\delta_{an}$ and MR in manganites is revealed. This correlation is 
rather apparent, but, surprisingly, was never mentioned in preceding 
papers.  Indeed, the temperature dependence $\delta_{an}(T)$ is entirely 
analogous to the MR temperature dependence (see insert in Fig. 1 or similar
graphs in the numerous CMR papers, for example, in Refs.\ 
\onlinecite{nagaev,ramirez,coey2}). The MR in the CMR manganites has some minimal 
magnitude at $T \ll T_c$, goes to maximal value near $T_c$, and approaches 
zero at $T > T_c$ in the same manner as the AMR parameter
$\delta_{an}(T)$ (Fig. 9). It should be noted that this type of temperature behavior 
of $\delta_{an}(T)$ and MR is a feature of manganite samples with fairly good 
crystal perfection only. In disordered doped manganites 
(for example, in polycrystalline or granular samples), the MR can rise with
decreasing temperature\cite{gupta96}. In that case, as shown in Ref.\ 
\onlinecite{ziese}, the parameter $\delta_{an}$ increases with 
decreasing temperature as well. 
\par
As was mentioned above,  the resistance $R$ of manganites in the ferromagnetic 
state is a function of magnetization.  The conductivity  
increases with enhancement of ferromagnetic order. This is the source of 
the huge negative MR in manganites. The MR magnitude is determined by the 
ability of an external magnetic field to increase  the magnetization. It 
is obvious that in good crystals at low temperature ($T \ll T_c$), when nearly 
all spins are already aligned by the exchange interaction, the ability to 
increase the magnetization in the magnetic field is minimal. For  
increasing temperature and, especially, at temperatures close to the Curie 
temperature $T_c$, the magnetic order becomes weaker (the magnetization goes 
down) due to thermal fluctuations. In this case the possibility to strengthen 
the magnetic order with an external magnetic field increases 
profoundly.  This is the reason 
for maximal MR magnitude near $T_c$. At last, above $T_c$, the spin arrangement 
becomes essentially random, the magnetization is zero, and, the MR is close to 
zero as well.  From the aforesaid, it might be assumed that the magnitude of the 
AMR effect in CMR manganites is determined by the same factors, as that of 
the MR, namely, by the possibility to increase the magnetic order by an  
applied magnetic field.  This feature must be important for further 
determination of the nature of the AMR effect in CMR manganites.   
\par
One further comment must be added. To see the AMR
effect properly, the magnetization should be parallel to the applied 
magnetic field.  That is, the magnitude of the field must be high enough to
saturate or rotate the magnetization to the selected direction. In other words,  
the applied field should be sufficient to overcome the different anisotropy 
energies of the film. Actually, the $\delta_{an}(H)$ behavior should reflect 
somehow the magnetization curve $M(H)$ in the direction perpendicular to the
film plane, and, hence, reflect the influence of the available anisotropy 
sources. As indicated above in Sec.\ \ref{magani}, the recorded $M(H)$ curves 
at low temperature ($T \ll T_c$) [Fig. 12 (a)] indicate that  some 
parts of the film have a substantial in-plane spontaneous magnetization, while 
other parts have a substantial out-of-plane one in a magnetic field which is 
very close to zero. We believe that the inflection in the $\delta_{an}(H)$ 
dependence (Fig.~8) at weak fields reflects this magnetization inhomogeneity 
in this film.

\subsection{Extrinsic MR anisoropy sources and their interplay with the 
AMR effect}

It is known that FM consists of magnetic domains which are regions of
spontaneous magnetization. Therefore, even in the absence of an external
magnetic field the electrons in FM feel the internal magnetic field 
$B = M_s$. For this reason, any measured resistance for a FM is, in fact, 
some measure of its MR as well. Apart from the influence of the intrinsic field,
there are additional specific mechanisms of electron scattering in
FM metals. Let us call them the ``magnetic'' mechanisms of electron scattering. 
These can give a considerable contribution to the resistivity and MR of FM 
metals\cite{vons,ross}. In the Matthiessen-rule approximation it is possible
to write down for the resistivity of FM metal
\begin{equation}
\rho (T, {\bf H}) = \rho_{nm}(T, {\bf H}) + \rho_{m}(T, \bf{H}), \eqnum{2}
\end{equation} 
where $\rho_{nm}$ is the ``non-magnetic'' part of the resistivity, which
is stemming from the usual electron-scattering mechanisms common to 
non-magnetic metals (scattering on impurities, phonons and so on), and  
$\rho_{m}(T, \bf{H})$  represents the ``magnetic'' part of the resistivity. 
\par
It has been long inferred that the behavior of $\rho_{m}(T, \bf{H})$ is some
reflection of the temperature and magnetic field dependences of the
magnetization. With a rise of magnetization at the transition into FM state,
$\rho_{m}$ drops sharply. In the FM state the ``magnetic'' resistivity is quite
small and, in the limit, goes to zero or to very low values for ideal 
magnetic order. The external magnetic field enhances the magnetic order, that 
leads to a decrease of resistivity. That is why FM metals are characterized by
a negative MR. Of course, at finite temperatures there are some thermal
disturbances of the long-range magnetic order (spin waves or magnons) which
can determine the power-law temperature dependence of $\rho_{m}$ at low
temperatures: $\rho_{m} \propto T^n$, where $n$ value depends on the specific
mechanism of disturbance.
\par
The known behavior of $\rho_{m}$ in FM metals\cite{vons,ross} indicates,
therefore,  that the influence of magnetic order (or a
magnetic lattice of spins) on electron transport is quite similar to that of 
the crystal lattice order. If the crystal lattice is ideal, the resistance is 
zero. The same is true (at least, to some significant
degree) in respect to magnetic order: for an ideal spin alignment the 
``magnetic'' resistivity may be thought to be equal to zero. Any deviations from
long-range magnetic order lead to electron scattering. Just as for
crystals, disorder may be static or dynamic disturbances (defects) in 
the spin lattice. It follows from all this that the ``magnetic'' part of the
resistivity, $\rho_{m}$, is a direct function of the magnetization, that is
\begin{equation}
\rho_{m} = f[M(T, {\bf H})], \eqnum{3}
\end{equation}  
The relevant experimental dependences (and its theoretical 
justifications) for mangnites can be found in Refs.\ \onlinecite{hund,font}. 
Lastly, the thermodynamic fluctuations of magnetic order should
be mentioned. They are especially strong near the Curie temperature, and, 
therefore, behavior of $d\rho/dT$ and sometimes that of resistivity (as in 
the case of manganites) has peculiarities at $T \approx T_c$.
\par
It should be noted in this place that the ``magnetic'' and ``non-magnetic'' 
contributions to the resistivity cannot be considered as entirely independent. 
The lattice defects and deviations from magnetic  order can be coupled 
rather strongly and be interdependent. For example, crystal lattice 
defects, such as grain boundaries, surface regions of film and others, 
induce disturbances in the magnetic order as well. On the other hand, 
the changes in the magnetic order, such as development of spontaneous 
magnetization at the paramagnetic-ferromagnetic
transition or moving and disappearence of domain walls in an external 
magnetic field, can cause a response of the crystal lattice (for example, 
changes in the elastic stresses and strains). 
\par
In explanation of the results of this study it is helpful, as a first step, to 
consider and keep in mind some known simple cases of 
the MR anisotropy in manganite films. An instructive example can be found in 
Ref.\ \onlinecite{eck} for films La$_{0.7}$Ca$_{0.3}$MnO$_{3}$, grown by 
molecular beam epitaxy on SrTiO$_3$ substrates. The films were between 50 
and 150 nm thick.  The authors of Ref.\ \onlinecite{eck} have studied  
$R(H)$ dependences for the cases where the magnetic field was
applied parallel or perpendicular to the film plane. For the perpendicular
field direction a positive MR was observed at low fields, which changed
to negative one at higher fields. The in-plane MR was only negative and 
depended on the angle, $\theta$, between transport current and field 
according to Eq. (1), which has been attributed to the AMR 
effect\cite{Mc,dahl}. The experimental dependences of $M(H)$ revealed 
anisotropy, which is favorable for the in-plane magnetization. Similar 
results were reported  also for Pr$_{0.67}$Sr$_{0.33}$MnO$_{3}$ 
films\cite{wangli} deposited on SrTiO$_3$ substrates.
\par
At first glance a positive MR in manganites appears to be quite impossible.
Really, the external magnetic field can only strengthen the long-range
magnetic order, and, therefore, should lead to decreasing resistance. 
Nevertheless, it turns out that a concurrence of surface (or shape) 
anisotropy and the AMR effect can cause the positive MR in a perpendicular 
field. A comprehensive explanation of this effect can be found in 
Ref.\ \onlinecite{eck}, so we will restrict ourself only to the main points,
which are necessary to understand the observations of this study. An essential 
prerequisite is that the MR of manganites be determined by the dependence of 
the magnetization on 
magnetic field. Assume now that field $H_z$ is applied perpendicular
to the film plane. At $H_z = 0$ the magnetization vector $\bf M$ has an in-plane
orientation (due to the influence of surface anisotropy and the demagnetization
energy). At low values of $H_z$ the applied field is actually perpendicular
to $\bf M$. For increasing $H_z$ the magnetization begins to rotate so that
a component $M_z$ appears which is perpendicular to the film 
plane\cite{dahl}.   As this proceeds the absolute magnitude of the magnetization 
remains unchanged up to the moment when $H_z$ reaches some field $H_s$ at which 
rotational saturation of the magnetization of the film in perpendicular 
direction takes place. The constancy of the absolute value of the 
magnetization during the rotation means that the ``magnetic''
part of resistivity, which depends on $M$, remains constant during the rotation 
as well.  In this case it should be expected that the MR would be zero up to 
the field $H_s$. Above this field a further increase in $H_z$ results in  
increasing $M$ and, therefore, in a  decreasing resistivity. However, instead 
of this, a positive MR was observed in low fields, and the negative one in 
higher fields. What is the reason for this behavior?
\par
The point is that the MR is affected also by the AMR effect (that is, by 
dependence of MR on the angle between the current and magnetization). At 
$H_z = 0$ the
magnetization vector lies in the film plane. In the explanation of 
Ref.\ \onlinecite{eck} it was implied that the current at $H_z = 0$ is parallel
to the magnetization. Maybe this is not rigorously correct, but it is not
so important, if an effective angle between $\bf M$ and $\bf J$ is fairly
small. For increasing $H_z$, the magnetization vector rotates, that is the 
angle between $\bf M$ and the current increases. This leads to the 
resistivity rise [see Eq. (1)] and is the cause of the positive MR.
The magnetization $\bf M$ becomes perpendicular to the film plane at 
$H_z = H_s$. At $H_z > H_s$ the magnetization begins to increase and this leads 
to the resistance decrease. As a result of this kind of concurrence the
dependence $R(H)$ with a maximum takes place. In the parallel field the
influence of AMR effect can be thought as absent, therefore only the decrease 
in resistance in magnetic field inherent for CMR manganites is observed.
\par
The results presented in this paper are quite different from these of 
Refs. \onlinecite{eck,wangli}. First of all, in the range of low magnetic
field (Figs. 4--6), for a perpendicular field, $H_{\bot}$, we found the 
negative MR in low field, before going to positive MR at higher fields 
(Fig. 4). At the same time, in a parallel field, $H_{\|}$, a positive MR is seen 
before becoming negative at higher fields (Fig. 5). This behavior is quite 
challenging and  puzzling. One of the most reasonable explanations 
for this is an inhomogeneous strain state of the film, that leads (due to the 
magnetoelastic interaction) to a  difference in the magnetic properties (for 
example, in directions of easy magnetization) in different parts of the film.
This turns out to be a basis for understanding of results of this study. 
The sources of internal strains and stresses in PLD films are quite common 
and sometimes inevitable due to the lattice film-substrate mismatch. It is just
these strains which can be the primary source of inhomogenenous magnetic state
of the film studied.
\par
It follows from the known studies\cite{wangli,nath,suzuki,odon} that 
magnetostriction in manganite films (at least with composition La-Ca-Mn-O 
and La-Sr-Mn-O) is positive. In this case, the magnetization orients parallel 
to the tensile stresses and perpendicular to the compressive stresses. 
For fairly smooth substrates, which make possible coherent epitaxy,
the in-plane film strain depends on the lattice film-substrate mismatch. 
If the mismatch is not too large, it can be expected that the in-plane 
lattice parameters will match those of a substrate and the out-of-plane 
parameter will be elastically modified according to the Poisson ratio. In this 
case, a biaxial strain is induced in the film plane which can be tensile or 
compressive 
depending on the ratio the substrate lattice parameters and those of the bulk 
target from which the film is deposited. This is true up to some critical 
thickness, above which misfit dislocations appear at the film-substrate 
interface to relax the strain. In that case, film lattice parameters 
become closer to those of bulk sample.  
\par
The above-described scenario is just a generally accepted model which
allows  some crude predictions to be made. XRD and other studies are required 
for  any specific film-substrate system to know exactly its strain state. 
Such studies have been  done in some  studies of the MR anisotropy in doped
manganite films.
For example, it was found that the La$_{0.7}$Ca$_{0.3}$MnO$_{3}$ films 
on SrTiO$_3$ substrate studied in Ref.\ \onlinecite{eck} have in-plane 
tensile strain\cite{odon}, that causes the in-plane magnetization and the 
appearence of positive MR in the field perpendicular to the film plane. In that
case the perfect matching of the in-plane lattice parameters of film and 
substrate was found. A different example (compressive strain 
Pr$_{0.67}$Sr$_{0.33}$MnO$_{3}$ films) is described in 
Ref.\ \onlinecite{wangli}. In this case the compressive strains cause the 
easy magnetization axis to be perpendicular to the film plane. It looks like 
that in the films, studied in Refs.\ \onlinecite{eck,wangli}, the strains, 
which are induced by the film-substrate interaction, were extended over the 
most of the film thickness. Consequently, the films in 
those studies can be considered as nearly (or to a great extent) 
homogeneous. This is especially true in respect to Ref.\ \onlinecite{wangli}
where the film thicknesses were between 7.5 and 15 nm.
\par 
For the La$_{1-x}$Ca$_{x}$MnO$_{3}$ ($x \approx 1/3$) film in this study,
grown on LaAlO$_3$, the substrate lattice parameters are less than those of 
the corresponding bulk sample. Therefore, a compressive biaxial strain in the 
film plane, and a corresponding tensile uniaxial strain in direction 
perpendicular to the film plane should be expected. This strain state has
been actually observed (see Sec.\ \ref{xray}). A perfect matching of the 
in-plane lattice parameters of film and substrate was not found, which is 
in agreement with the study of La-Ca-Mn-O films on LaAlO$_3$ in 
Ref.\ \onlinecite{nath}. This is because of partial relaxation of the strain 
emposed by the substrate. Additionally, the XRD data indicate
that the crystall structure of the film is inhomogenenous. It consists of 
the regions with different crystallographic orientations (see 
Sec. \ \ref{xray}). The origin of the inhomogeneous structure is probably 
connected with the twin structure of the substrate. As a result, in 
both the in-plane and out-of-plane directions tensile and compressive
regions can be found. 
\par
The inhomogeneous film structure reveals itself in the behavior of $M(H)$ 
curves [see (Fig. 12) and the discussion in Sec.\ \ref{magani}] as well as in 
the magnetic field dependence of the AMR parameter (Fig. 8) (in the last 
case we mean the inflection in the $\delta_{an}(H)$ curve which should not 
occur for a  homogeneous system). The film inhomogeneity is a key to 
understanding the observed MR anisotropy (Fig. 4 and
5) as well. It can be assumed that the film is some mixture of regions with 
different strains. The size of the regions (according to the XRD data) is 
comparable with the film thickness. The regions with the in-plane compressive
strains favor the out-of-plane magnetization in zero field; whereas, the ones 
with the in-plane tensile strains favor the in-plane magnetization. In the 
following text we can conveniently speak about the ``compressive''
and ``tensile'' parts of the film. In spite of the fact that some part of 
the film is prone to the out-of-plane magnetization, the influence of the 
shape anisotropy may cause the total film magnetization to be in the film 
plane in zero magnetic field or to be canted, as was found, for example, in 
La-Sr-Mn-O films under compressive stress\cite{kwon}. In any case, however, 
it is possible to understand the appearence of a negative MR in low perpendicular 
fields, $H_{\bot}$, (Fig. 4) for an inhomogenenous  film of this type .
The ``compressive'' part of the film can be magnetized in the perpendicular 
direction at very low magnitude of $H_{\bot}$ up to the rotation saturation 
value, and after this the absolute value of magnetization begins to increase 
with increasing
magnetic field. This leads to a resistance decrease, that is, to a
negative MR. At the same time, the magnetization of the ``tensile'' part of
the film is in the film plane at zero field and cannot be rotated so easily
in the field direction. Considerably higher fields are needed for it.
Therefore, at higher fields the explanantion given in Ref.\ \onlinecite{eck},
which takes into accont the AMR effect, is quite applicable to justify
the positive MR at higher fields. All these combined effects can produce the 
observed rather complicated $R(H_{\bot})$ dependence (Fig. 4). 
\par
Let us turn now to Fig. 5 which represents the behavior of the resistance in
a magnetic field, $H_{\|}$, parallel to the plane of the film. In this case the
resistance goes up initially for increasing field, but then goes down in 
higher field. That is, the change in the MR sign (from positive to negative) 
takes place. To understand this, it should be recalled that both, 
$R(H_{\bot})$ and $R(H_{\|})$, dependencies (Figs. 4 and 5) were registered 
in magnetic fields perpendicular to the current. At low magnitudes of 
$H_{\|}$, however, the film magnetization is definitely not perpendicular to 
the current. At least in the ``tensile'' regions of the film with the 
magnetization 
easy axis parallel to the film plane, the magnetization is by no means 
perpendicular to the current. The vectors $\bf M$ in these regions should
have some spread in directions due to the demagnetization energy. Only for 
increasing $H_{\|}$ do these vectors 
become strictly perpendicular to the current. Since MR is maximal when 
the magnetization is perpendicular to the current due to the AMR effect 
(Fig. 7), the in-plane rotation of magnetization in low fields $H_{\|}$ 
leads to an increase in resistance. This explains  the positive MR for
small parallel fields. After aligning the spins in the ``tensile'' regions
parallel to magnetic field, the magnetization begin to increase which causes
as usual the resistance decrease. It should be mentioned that a similar effect 
was observed in Ref.\ \onlinecite{li} for a 
La$_{0.67}$Sr$_{0.33}$MnO$_3$ film for magnetic fields applied in the film 
plane. They found a positive MR for $H\bot J$ as opposed to the negative 
one with $H\| J$. $R(H)$ curves were measured, however, only in low fields 
($H < 1.5$~kOe), which thus excludes a comparison with the data of this paper 
(registered in fields up to about 20 kOe).  
\par
In fields parallel to both, the current $J$ and the film plane, only 
the negative MR is found in this study (Fig. 6). This is quite expected, 
since there are no mechanisms for positive MR in this case.

\section{Conclusion}
The MR behavior in the PLD manganite films can differ dramatically from that of 
bulk samples. This is due to a film-substrate interaction, which determines the 
structural and magnetic state of the films. Some film-substrate combinations 
can lead to rather complicated and puzzling MR behavior for different 
directions of the magnetic field relative to the film plane and the transport 
current. To understand such cases properly, one needs to have enough data about 
the structural and magnetic properties of the films. The MR and MR anisotropy 
of such
film systems depends on the existence of preferential directions of
magnetization (due to the strains arising from the lattice film-substrate
mismatch and other sources), and from the AMR effect. We have presented in
this paper an example of complicated behavior of the low-field MR and MR
anisotropy for La$_{1-x}$Ca$_{x}$MnO$_{3}$ films on LaAlO$_{3}$
substrates, and demonstrated how it can be understood.
\par
Based on the results of this study together with the known results of other 
authors, we have indicated fairly conlusively  (and for the first time) that a 
clear correlation exists between the magnitudes of the AMR effect and MR in 
manganites. This suggests that the AMR effect in manganites is determined by the
ability of the magnetization to increase in an external magnetic field. This 
important correlation can be helpful in
further disclosing the nature of the AMR effect in CMR manganites.

\acknowledgments
The authors are indebted to Dr. N. V. Dalakova for help in measuring $R(T)$ 
dependence in zero magnetic field and Dr. J. H. Ross for his efforts in 
measuring of film thickness by AFM. Support at
TAMU was provided by the Robert A. Welch Foundation (grant A-0514) 
and THECB ARP 010366-003.  BIB and DGN acknowledge support
by NATO Scientific Division (Collaborative Research Grant No. 972112).

\begin{figure}[htb]
\caption{Temperature dependence of the resistance of the
investigated  film. The insert shows the temperature dependence of
$\delta_{H} = [R(0) - R(H)]/R(0)$ at $H=10$~kOe (filled triangles) and
$H=16$~kOe (filled circles). Field $H$ was perpendicular to the film plane.} 
\label{fig1}
\end{figure}

\begin{figure}[htb]
\caption{Temperature dependences of magnetization at external field 
$H = 200$~Oe for field directions parallell (a) and perpendicular (b) to
the film plane. The dependences were recorded on heating, after the film had
been cooled down from the room temperature to $T=4$~K in zero field. 
The magnetization is normalized to the saturation value, $M_0$, determined 
from $M(H)$ measurements at $T = 4$~K.} 
\label{fig2}
\end{figure}

\begin{figure}[htb]

\caption{Temperature dependences ac susceptibility  in ac magnetic field 
$H_{ac}=1$~Oe at frequency 125 Hz for field directions parallel (a) and 
perpendicular (b) to the film plane.} 
\label{fig3}
\end{figure}

\begin{figure}[htb]
\caption{Magnetoresistance at different temperatures  for the field
$H_{\bot}$ applied perpendicular to the film plane.} 
\label{fig4}
\end{figure}

\begin{figure}[htb]
\caption{Magnetoresistance at different temperatures  for the field
$H_{\|}$ applied parallel to the film plane.} 
\label{fig5}
\end{figure} 

\begin{figure}[htb]
\caption{Magnetoresistance at $T = 4.2$~K for different orientations of the
magnetic the field relative to the film plane and current 
$J$.}
\label{fig6}
\end{figure} 

\begin{figure}[htb]
\caption{The angular dependence of resistance in magnetic field $H=16$~kOe 
at $T=4.2$~K (solid circles connected by doted line). Solid line curve 
presents an equation
$R(\theta)/R(0) = 1 + \delta_{an}\sin^{2}\theta$ with $\delta_{an} = 
2.14\times 10^{-3}$. The $\theta$ is the angle between the field and current 
directions. The field has been rotated in a plane which is perpendicular to 
the film plane and parallel to the current. In such geometry, 
the position, where $\theta =0$, corresponds to $H$ parallel to the film 
plane, and the position, where $\theta =90^{\circ}$ corresponds to 
$H$ perpendicular to the film plane.}
\label{fig7}
\end{figure} 

\begin{figure}[htb]
\caption{The magnetic field dependence of the AMR parameter $\delta_{an}$ 
in  Eq. (1) at $T=4.2$~K. Solid line presents a B-spline fitting.} 
\label{fig8}
\end{figure} 

\begin{figure}[htb]
\caption{Temperature dependence of the AMR parameter $\delta_{an}$ 
in  Eq. (1) at $H=16$~kOe. The solid line presents a B-spline fitting.} 
\label{fig9}
\end{figure} 

\begin{figure}[htb]
\caption{The dependence  of $[(R_{\bot}/R_{\|})- 1]$ on magnetic field at 
$T=4.2$~K.
$R_{\bot}$ and $R_{\|}$ are resistances, recorded in magnetic fields 
perpendicular  and parallel to the film plane, respectively. In both 
cases the fields were  perpendicular to the transport current $J$.} 
\label{fig10}
\end{figure} 

\begin{figure}[htb]
\caption{The ratio of MR in magnetic fields parallel and perpendicular to
the film plane as a function of temperature at $H = 15$~kOe. 
The magnetic fields were perpendicular to the transport current $J$ for
both field directions.} 
\label{fig11}
\end{figure} 

\begin{figure}[htb]
\caption{The $M(H)$ curves at $T=4$~K (a) and $T=100$~K (b) for the film 
studied. The curves formed by filled circles and triangles are relevant
to magnetic fields parallel and perpendicular to the film plane, 
correspondingly.} 
\label{fig12}
\end{figure}

\end{multicols}

\begin{references}
\bibitem[*]{byline}  E-mail: belevtsev@ilt.kharkov.ua
\bibitem[\dag]{byline}  E-mail: naugle@physics.tamu.edu 

\bibitem{helm93}R. von Helmont, J. Wecker, B. Holzapfel, 
L. Schultz, and K. Samwer, Phys. Rev. Lett. {\bf
71,} 2331 (1993).

\bibitem{jin94}S. Jin, T. H. Tiefle, M. McCormack, R. A. Fastnacht,
R. Ramesh, and L. H. Chen, Science {\bf 264,} 413 (1994); 
S. Jin, M. McCormack, T. H. Tiefel, and R. Ramesh, J. Appl. Phys. {\bf 76,}
6929 (1994).  

\bibitem{hund}M. F. Hundley, M. Hawley, R. H. Heffner, Q. X. Jia, 
J. J. Neumeier, J. Tesmer, J.~D.~Thompson, and X. D. Wu, 
Appl. Phys. Lett. {\bf 67,} 860 (1995).

\bibitem{gupta96}A. Gupta, G. Q. Gong, Gang Xiao, P. R. Duncombe, 
P. Lecoeur, P. Trouilloud, Y.~Y.~Wang, V. P. Dravid,  and J. Z. Sun, 
Phys. Rev. B {\bf 54,} R15629 (1996).

\bibitem{zener}C. Zener, Phys. Rev. {\bf 82,} 403 (1951).

\bibitem{and55}P. W. Anderson and H. Hasegawa, 
Phys. Rev. {\bf 100,} 675 (1955).

\bibitem{gennes}P. G. de Gennes, Phys. Rev. {\bf 118,} 141 (1960).

\bibitem{nagaev}E. L. Nagaev, Physics--Uspekhi {\bf 39,} 781 (1996); 
Phys. Rev. B {\bf 54,} 16608 (1996).

\bibitem{ramirez}A. P. Ramirez, J. Phys., Cond. Matter {\bf 9,} 8171 (1997).

\bibitem{khom}D. I. Khomskii and G. A. Sawatzky, Solid State Commun. 
{\bf 102,} 87 (1998).

\bibitem{coey2}J. M. D. Coey, M. Viret, and S. von Molnar, Adv. Phys. 
{\bf 48,} 167 (1999). 

\bibitem{dagot}A. Moreo, S. Yunoki, E. Dagotto, Science {\bf 283,} 2034
(1999).

\bibitem{saitoh}T. Saitoh, A. E. Bocquet, T. Mizokawa, H. Namatame, 
A. Fujimori, M. Abbade, Y. Takeda, and M. Takano, Phys. Rev. B {\bf
51,} 13942 (1995).

\bibitem{ju}H. L. Ju, H.-C. Sohn, Kannan M. Krishnan, Phys. Rev. Lett. 
{\bf 79,} 3230 (1997).

\bibitem{croft}M. Croft, D. Sills, M. Greenblatt, C. Lee, S.-W. 
Cheong, K.V. Ramanujachary, and D.~Tran, Phys. Rev. B {\bf 55,} 8726 (1997).

\bibitem{pickett}W. E. Pickett and D. J. Singh, Phys. Rev. B {\bf 53,} 1146
(1996).

\bibitem{mahen}R. Mahendiran, S. K. Tiwary, A. K. Raychaudhuri,  
T. V. Ramakrishnan, R Manesh, N.~Rangavittal, and C. N. R. Rao, 
Phys. Rev. B {\bf 53,} 3348 (1996).

\bibitem{font}J. Fontcuberta, B. Martinez, A. Seffar, S. Pi\~{n}ol, 
J. L. Garcia-Mu\~{n}oz, and X. Obradors, Phys. Rev. Lett. {\bf 76,}
1122 (1996).

\bibitem{eck}J. N. Eckstein, I. Bozovic, J. O'Donnell, M. Onellion, 
and M.R. Rzchowski, Appl. Phys. Lett. {\bf 69}, 1312 (1996).

\bibitem{li}X. W. Li, A. Gupta, Gang Xiao, and G. Q. Gong,
Appl. Phys. Lett. {\bf 71}, 1124 (1997). 

\bibitem{wangli}H. S. Wang, Qi Li, Kai Liu, and C. L. Chien, 
Appl. Phys. Lett. {\bf 74}, 2212 (1999). 

\bibitem{Mc}T. R. McGuire and R. I. Potter, IEEE Trans. Magn. 
{\bf Mag-11,} 1018 (1975).

\bibitem{dahl}E. Dan Dahlberg, Kevin Riggs, G. A. Prinz, J. Appl. Phys.
{\bf 63,} 4270 (1988).

\bibitem{bel1}B. I. Belevtsev, V. B. Krasovitsky, D. G. Naugle, K. D. D.
Rathnayaka, A. Parasiris, Physica B {\bf 284,} Part 2, 1988 (2000). 

\bibitem{pandey1}S. R. Surthi, S. Bhat, R. K. Pandey, K. D. D. Rathnayaka,
A. Parasiris, A. C. Du Mar and D. G. Naugle, in {\it Integrated Thin Films 
and Applications} (Ceramic Transactions, vol. 86), edited by R. K. Pandey, 
David E. Witter, Usha Varshney (The American Ceramic Society, Westerville,
Ohio, 1998) pp. 109-118.  

\bibitem{bel2}In Ref.\ \onlinecite{bel1} we have estimated the film thickness
to be $\approx 120$~nm. After that we have measured the film 
thickness with AFM and find out the mean thickness is about 80 nm, but 
the surface is rather rough, so the highest points can reach about 100 nm. 

\bibitem{rom}M. A. Rom, I. N. Chukanova, Functional Materials {\bf 6,} 
915 (1999).

\bibitem{bond}W. L. Bond, Acta Crystallogr., {\bf 13,} 814 (1960).

\bibitem{surf}S. Bueble, K. Knorr, E. Brecht, W. W. Schmahl, Surf. Sci. {\bf 400,}
345 (1998).

\bibitem{bulk}R. B. Praus, G. M. Gross, F. S. Razavi, and H.-U. Habermeier,
J. Magn. Magn. Mater {\bf 211,} 41 (2000); R. Laiho, K. G. Lisunov, 
E. L\"{a}hderanta, P. Petrenko, V. N. Stamov, and V. S. Zakhvalinskii, 
{\it ibid.} {\bf 213,} 271 (2000).

\bibitem{koo} T. Y. Koo, S. H. Park, K.-B. Lee, and Y. H. Jeong, Appl. Phys.
Lett. {\bf 71,} 977 (1997).

\bibitem{nath}T. K. Nath, R. A. Rao, D. Lavric, C. B. Eom, L. Wu, and F. Tsui,
Appl. Phys. Lett. {\bf 74,} 1615 (1999); R. A. Rao, D. Lavric, T. K. Nath,
C. B. Eom, L. Wu, and F. Tsui, J. Appl. Phys. {\bf 85,} 4794 (1999). 

\bibitem{ara}F. M. Araujo-Moreira, M. Rajeswari, A. Goyal, K. Ghosh,
V. Smolyaninova, T. Venkatesan, C. J. Lobb, and R. L. Greene, Appl. Phys.
Lett. {\bf 73,} 3456 (1998). 

\bibitem{schif}R. Schiffer, A. P. Ramirez, W. Bao, and S.-W. Cheong,
Phys. Rev. Lett. {\bf 75,} 3336 (1995). 

\bibitem{neu}J. J. Neumeier. M. F. Hundley, J. M. Thompson, and 
R. H. Heffner, Phys. Rev. B {\bf 52,} R7006 (1995). 

\bibitem{shre}R. Shreekala, M. Rajeswari, R. C. Srivastava, K. Ghosh, A. 
Goyal, V. V. Srinivasu, S. E. Lofland, S. M. Bhagat, M. Downes, R. P. Sarma, 
S. B. Ogale, R. L. Greene, R. Ramesh, T. Venkatesan, R. A. Rao, and C. B. Eom,
Appl. Phys. Lett. {\bf 74,} 1886 (1999).

\bibitem{millis}A. J. Millis, A. Goyal, M. Rajeswari, K. Ghosh, R. Shreekala,
R. L. Greene, R. Ramesh, and T. Venkatesan (unpublished).

\bibitem{ziese}M. Ziese and S. P. Sena, J. Phys.: Condens Matter {\bf 10,} 
2727 (1998); M. Ziese, Phys. Rev. B {\bf 62,} 1044 (2000).

\bibitem{amaral}V. S. Amaral, A. A. C. S. Louren\c{c}o, J. P. Ara\'{u}jo, 
A. M. Pereira, J. B. Sousa, P. B. Tavares, J. M. Vieira, E. Alves, 
M. F. da Silva, and J. C. Soares, J. Appl. Phys. {\bf 87,} 5570 (2000).

\bibitem{vons}S. V. Vonsovsky, {\it Magnetism} (Nauka, Moscow, 1971).

\bibitem{stampe}P. A. Stampe, H. P. Kunkel, Z. Wang, and G. Williams, Phys. Rev.
B {\bf 52,} 335 (1995).

\bibitem{ross}P. L. Rossiter, {\it The electrical resistivity of
metals and alloys} (Cambridge University Press, Cambridge, 1987).

\bibitem{suzuki}Y. Suzuki, H. T. Hwang, S-W. Cheong, and R. B. van Dover,
Appl. Phys. Lett. {\bf 71,} 140 (1997). 

\bibitem{odon}O'Donnel, M. S. Rzchowski, J. N. Eckstein, and I. Bozovic,
Appl. Phys. Lett. {\bf 72,} 1775 (1998)

\bibitem{kwon}C. Kwon, M. C. Robson, K.-C. Kim, J. Y. Gu, S. E. Lofland,
M. S. Bhagat, Z. Trajanovic, M. Rajeswari, T. Venkatesan, A. R. Kratz,
R. D. Gomez, and R. Ramesh, J. Magn. Magn. Mater. {\bf 172,} 229 (1997).
\end{references}
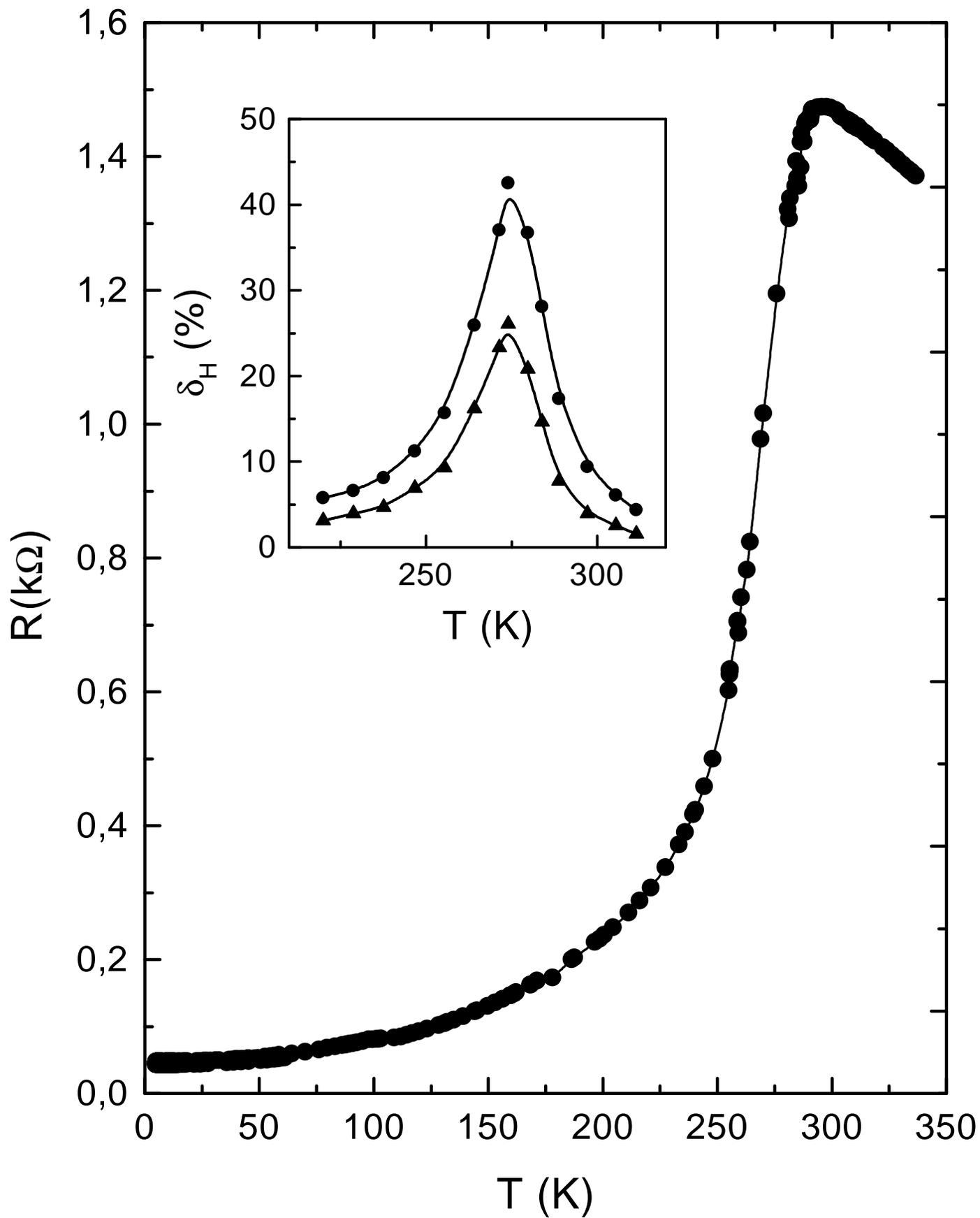
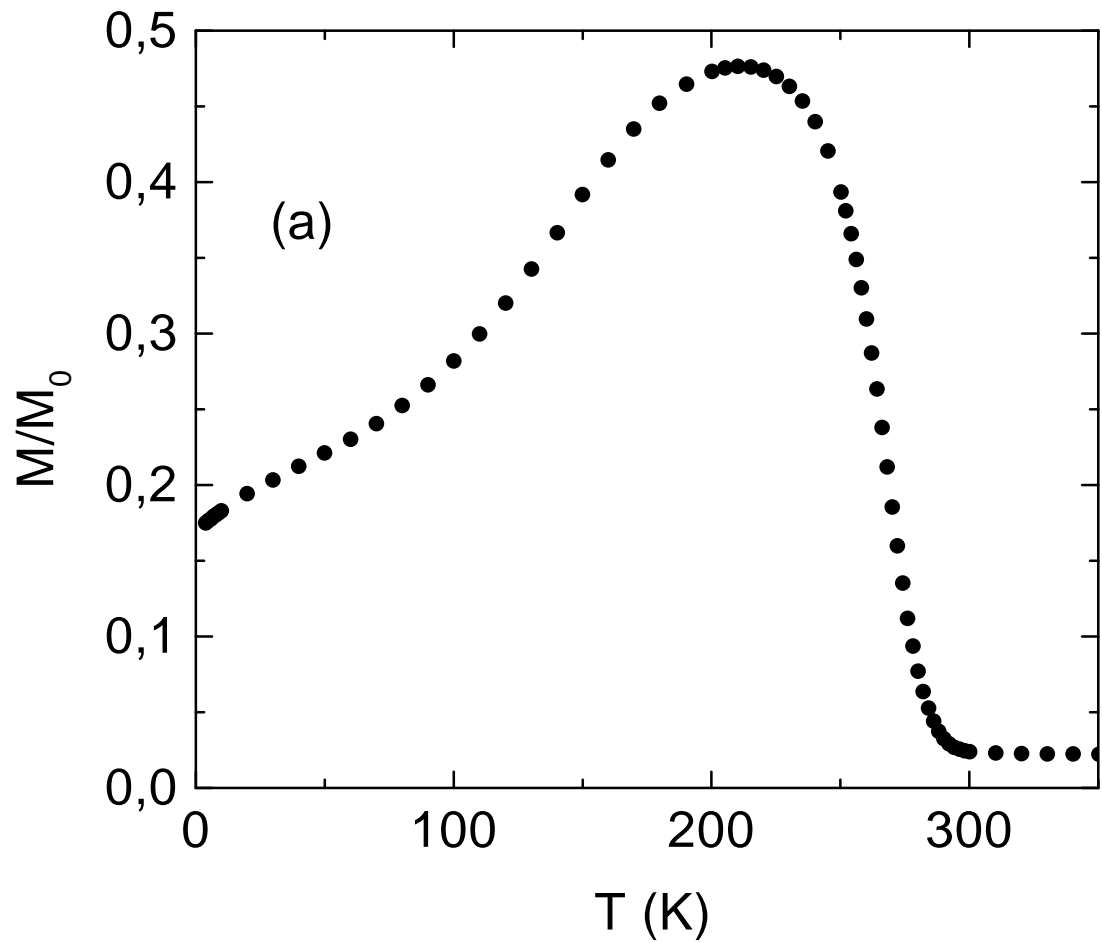
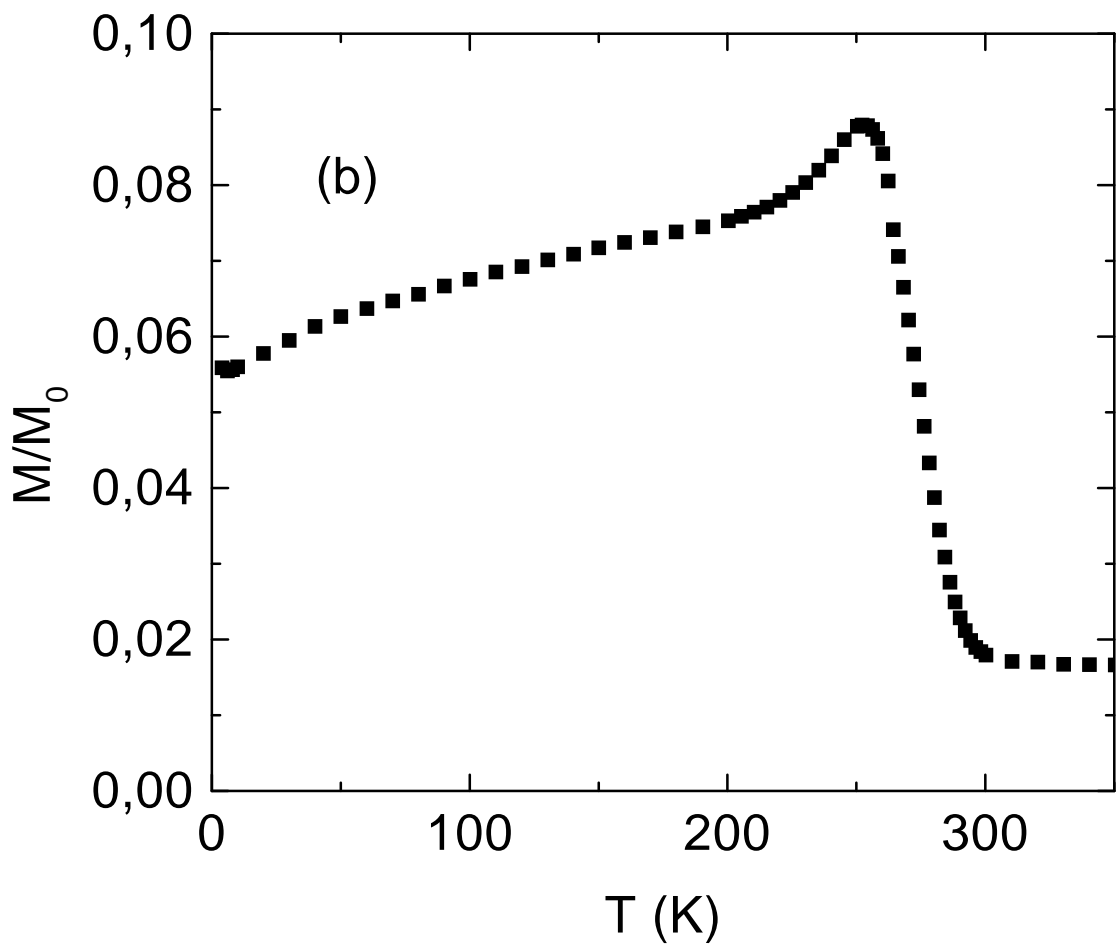
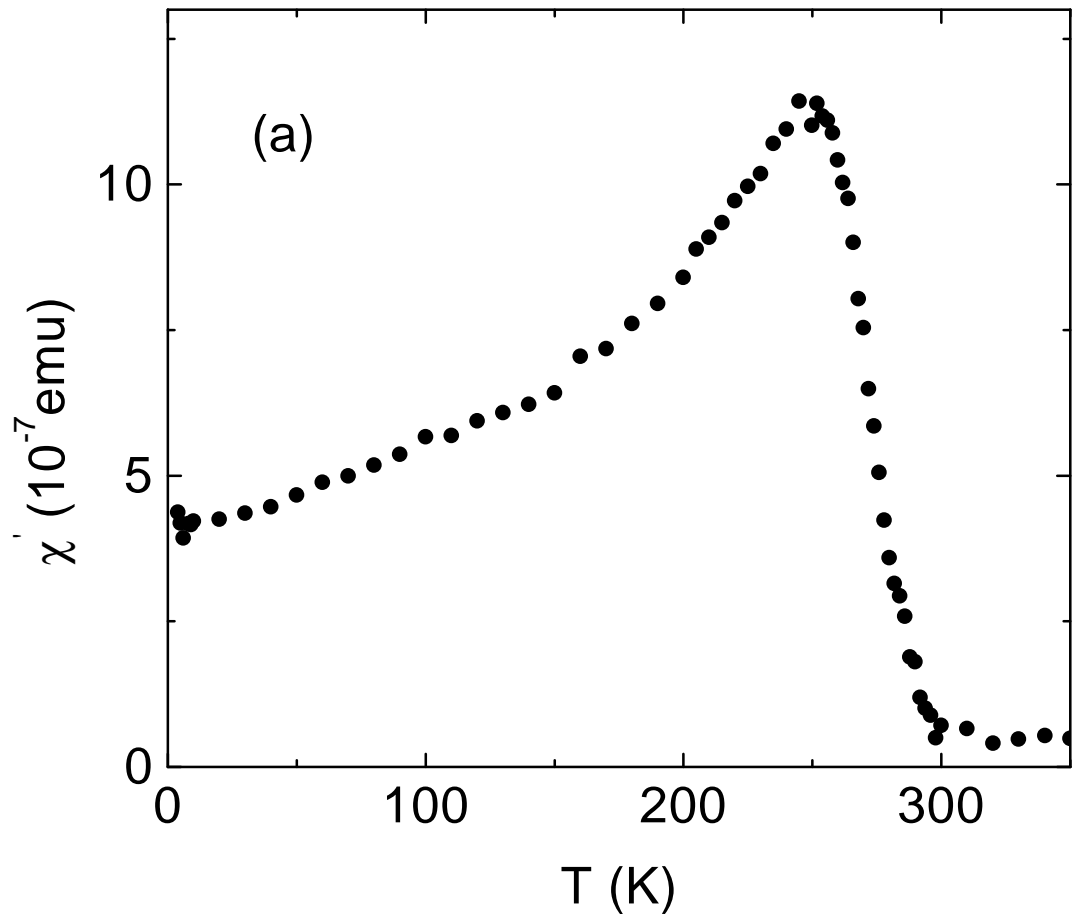
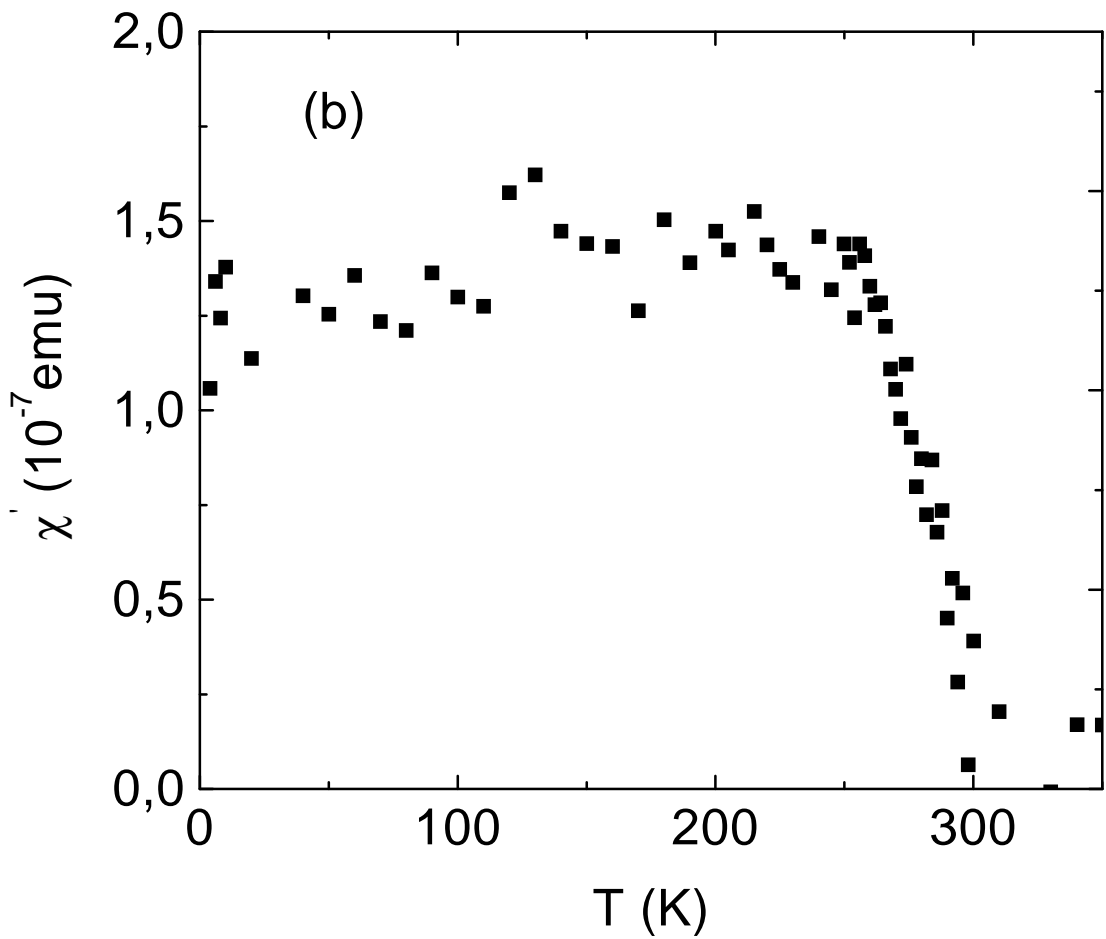
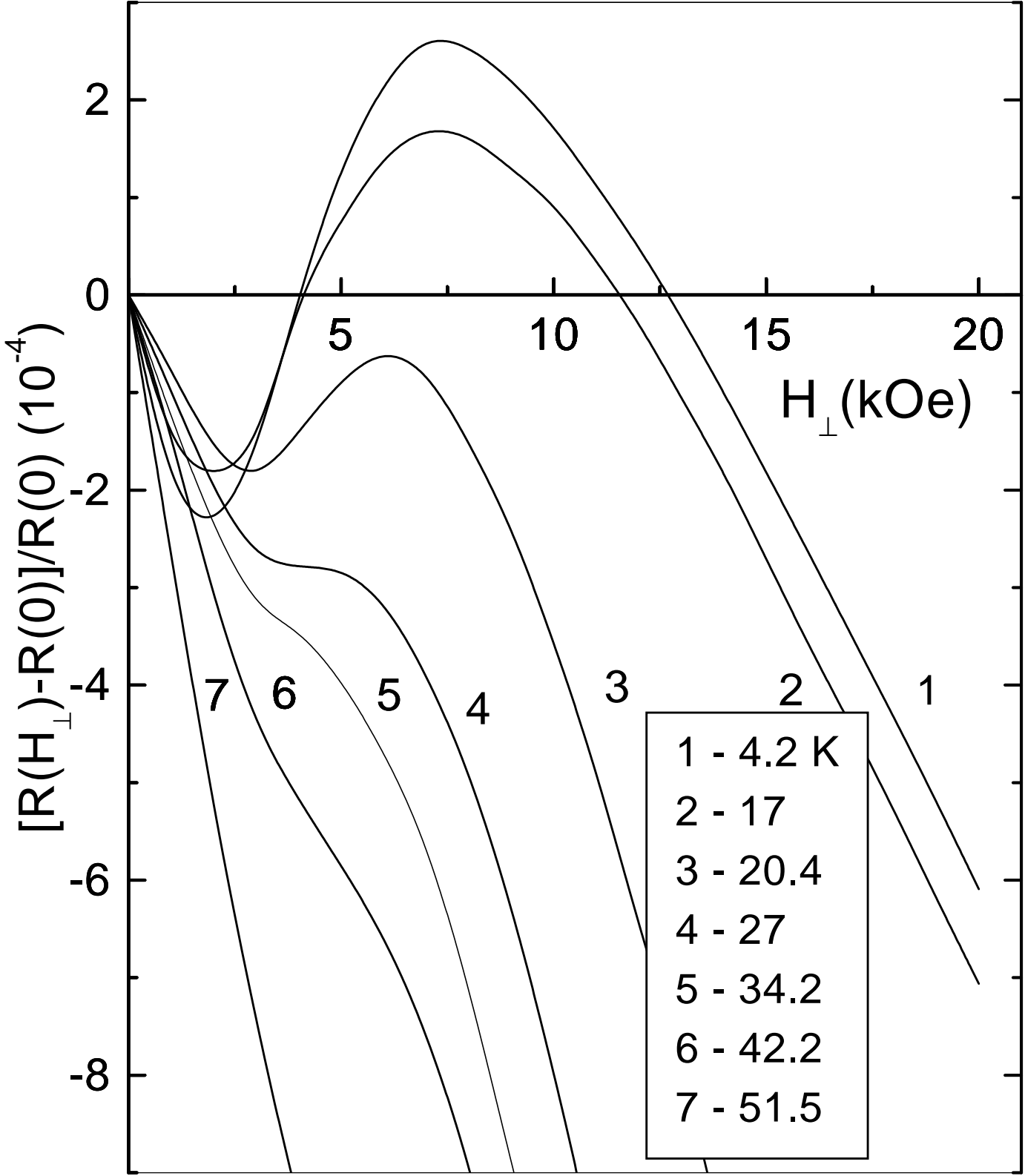
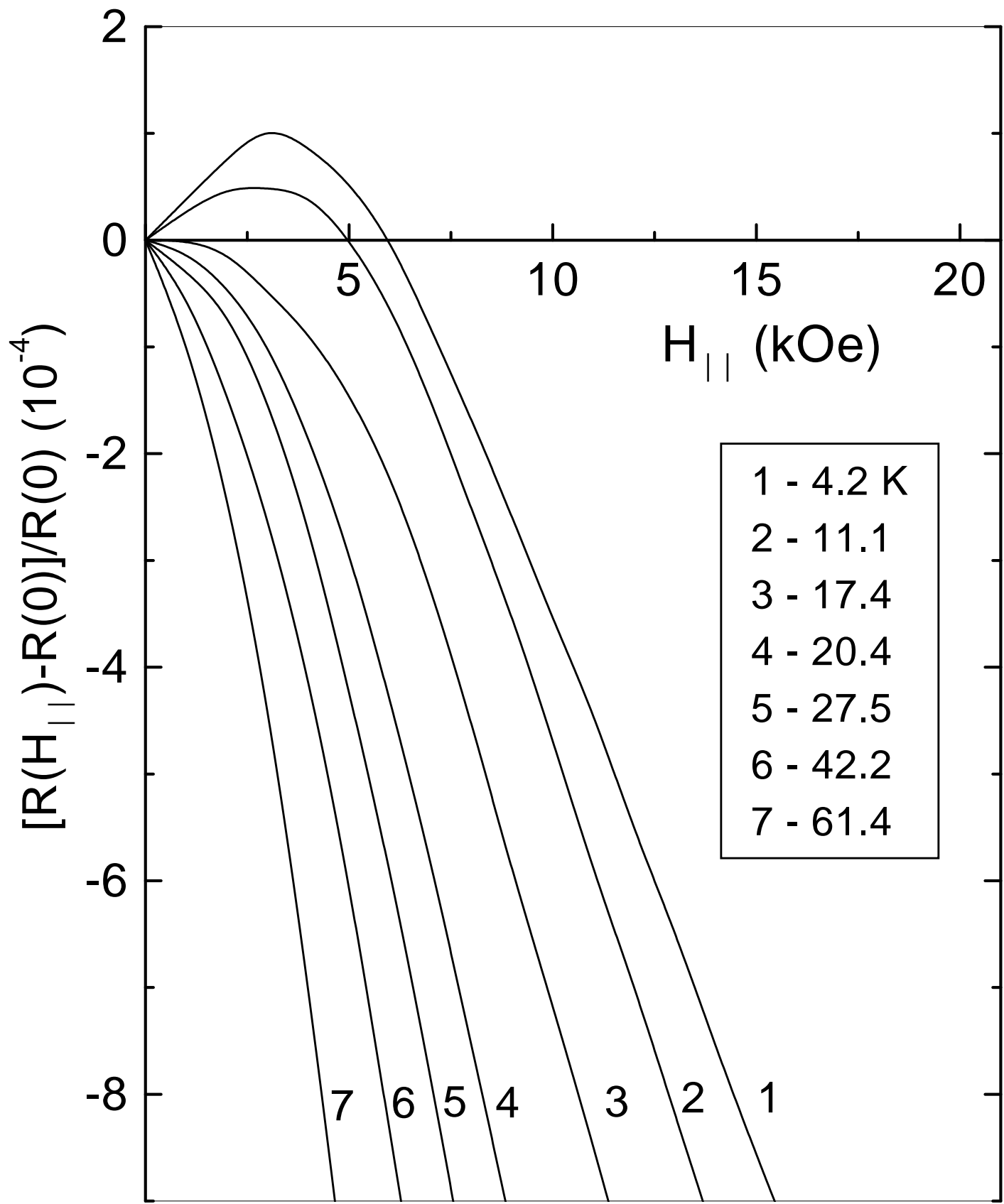
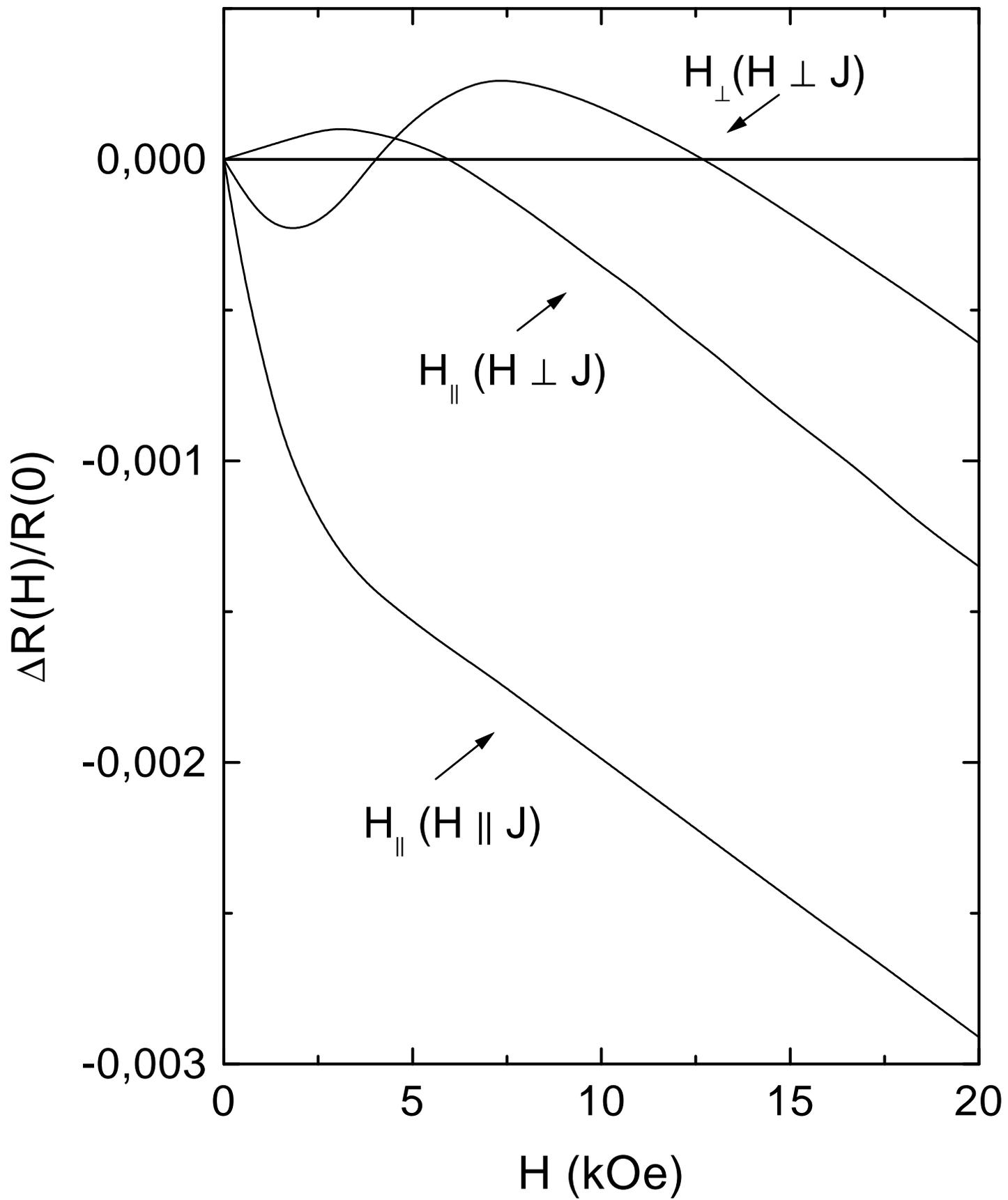
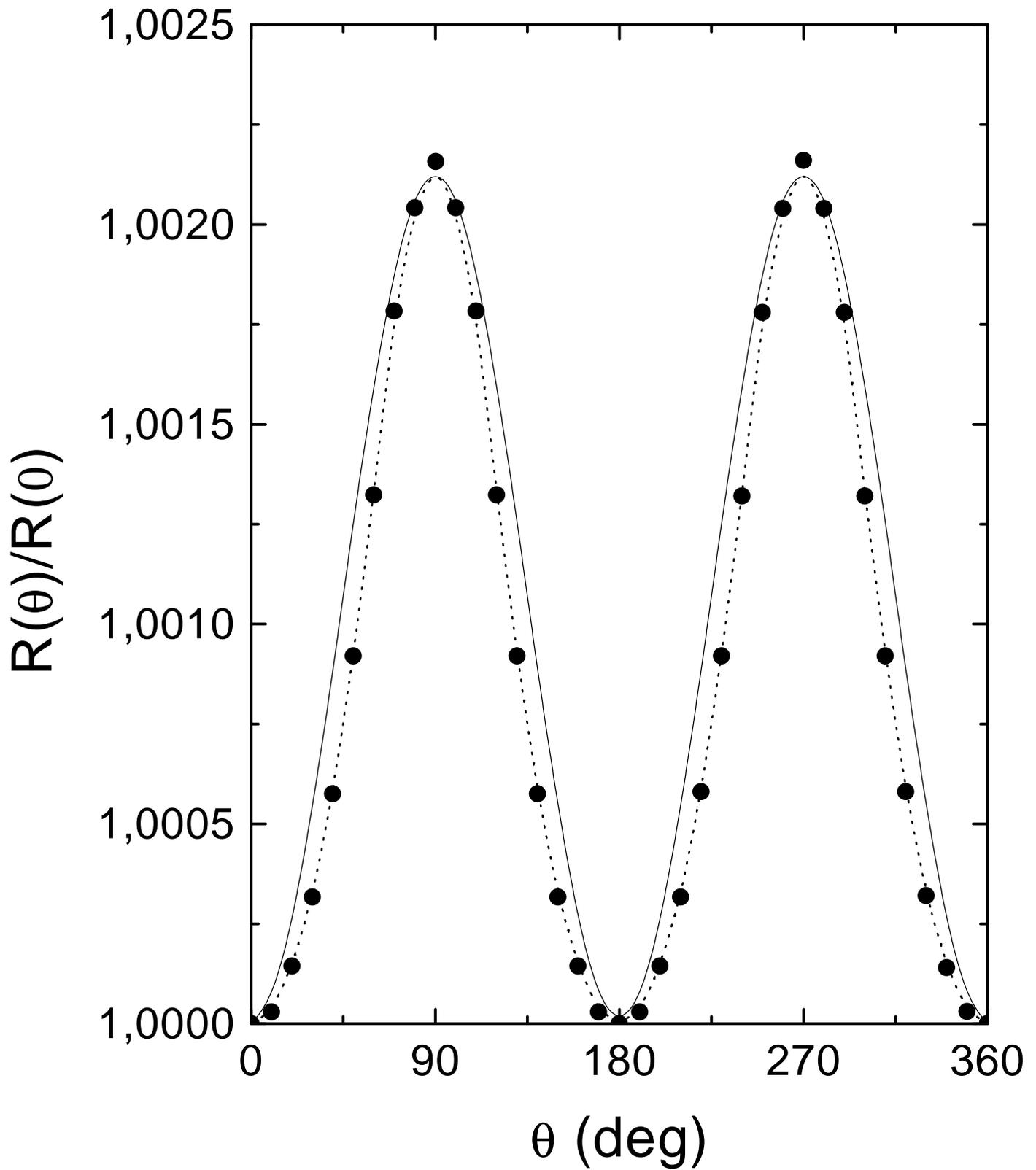
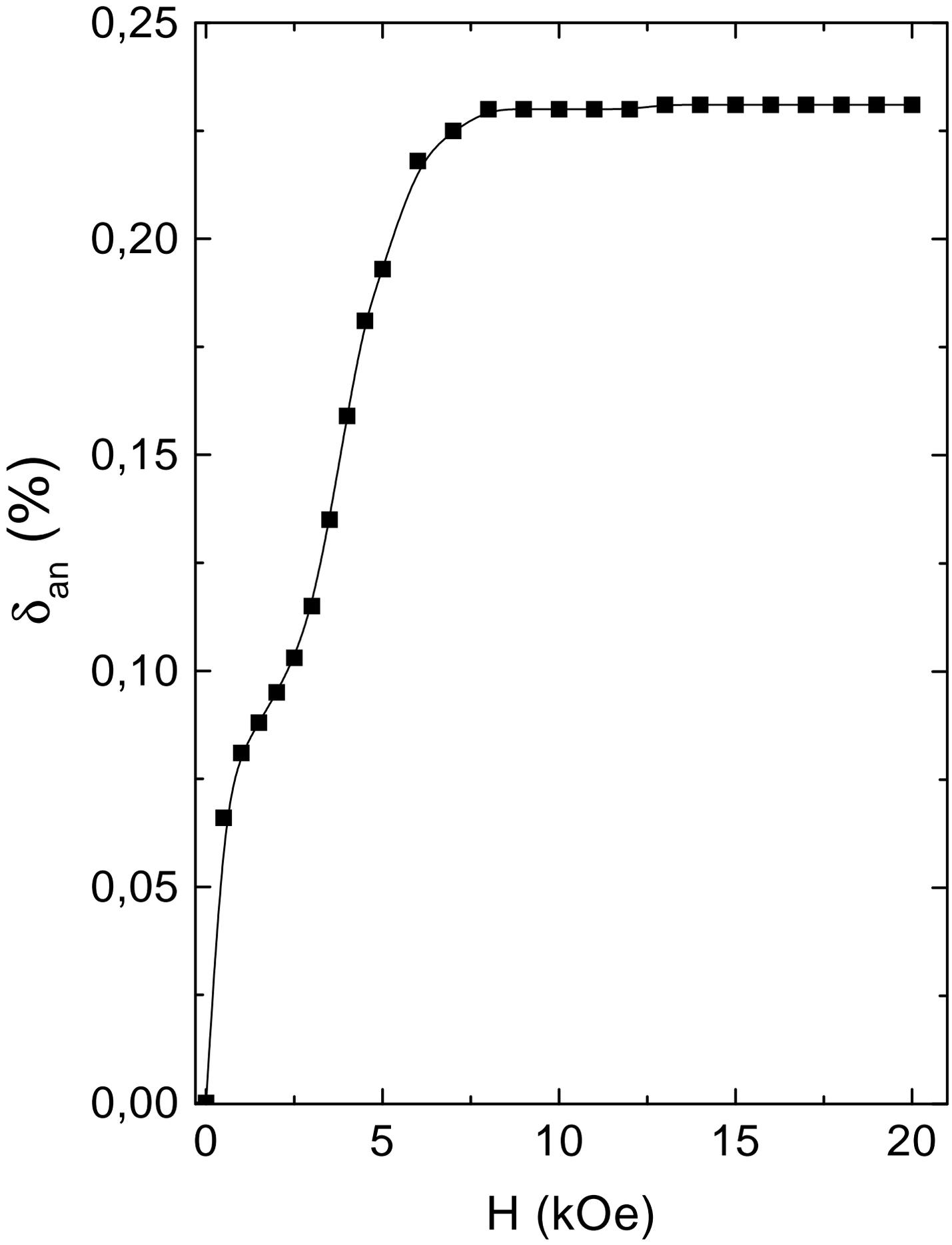
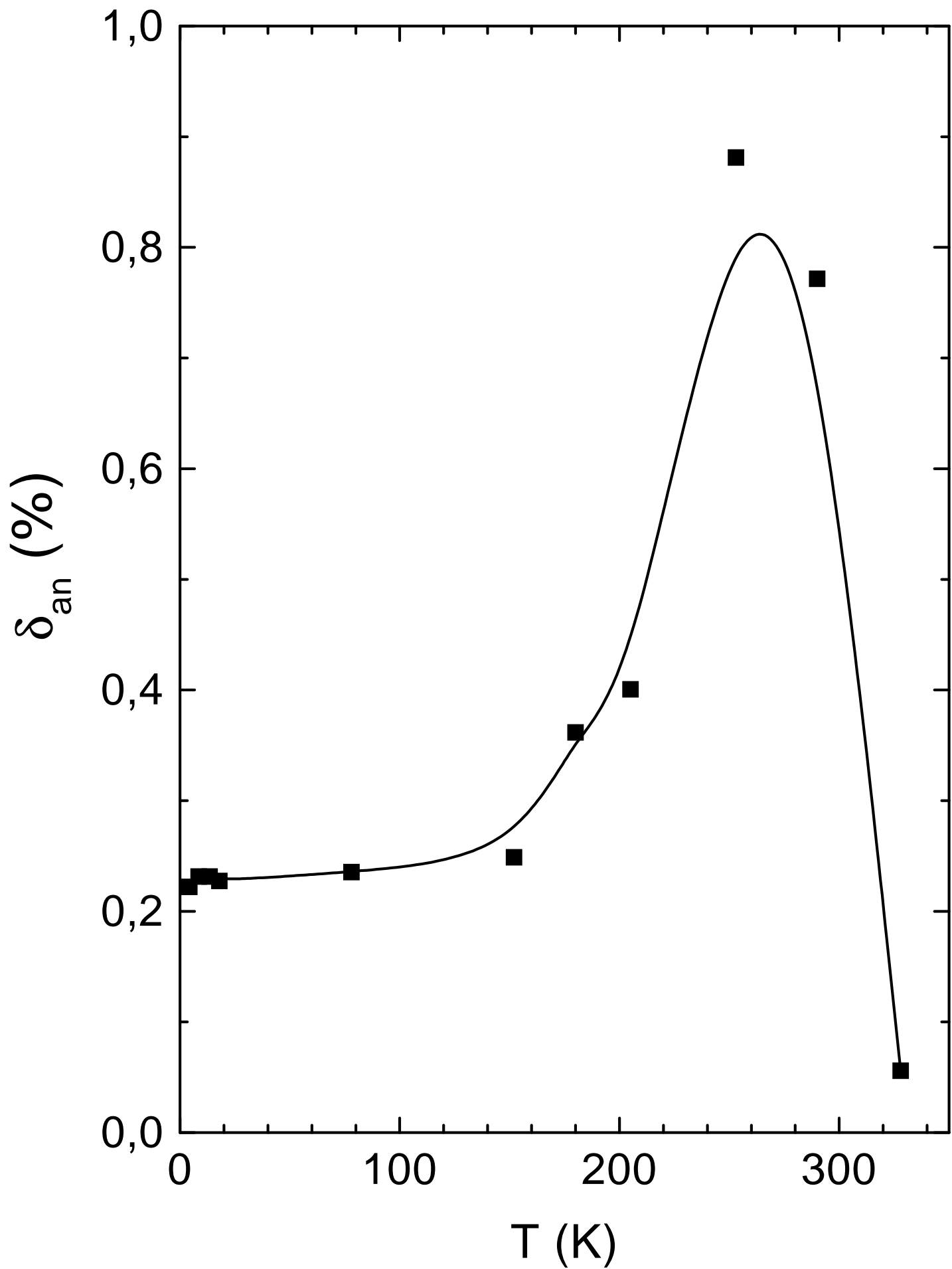
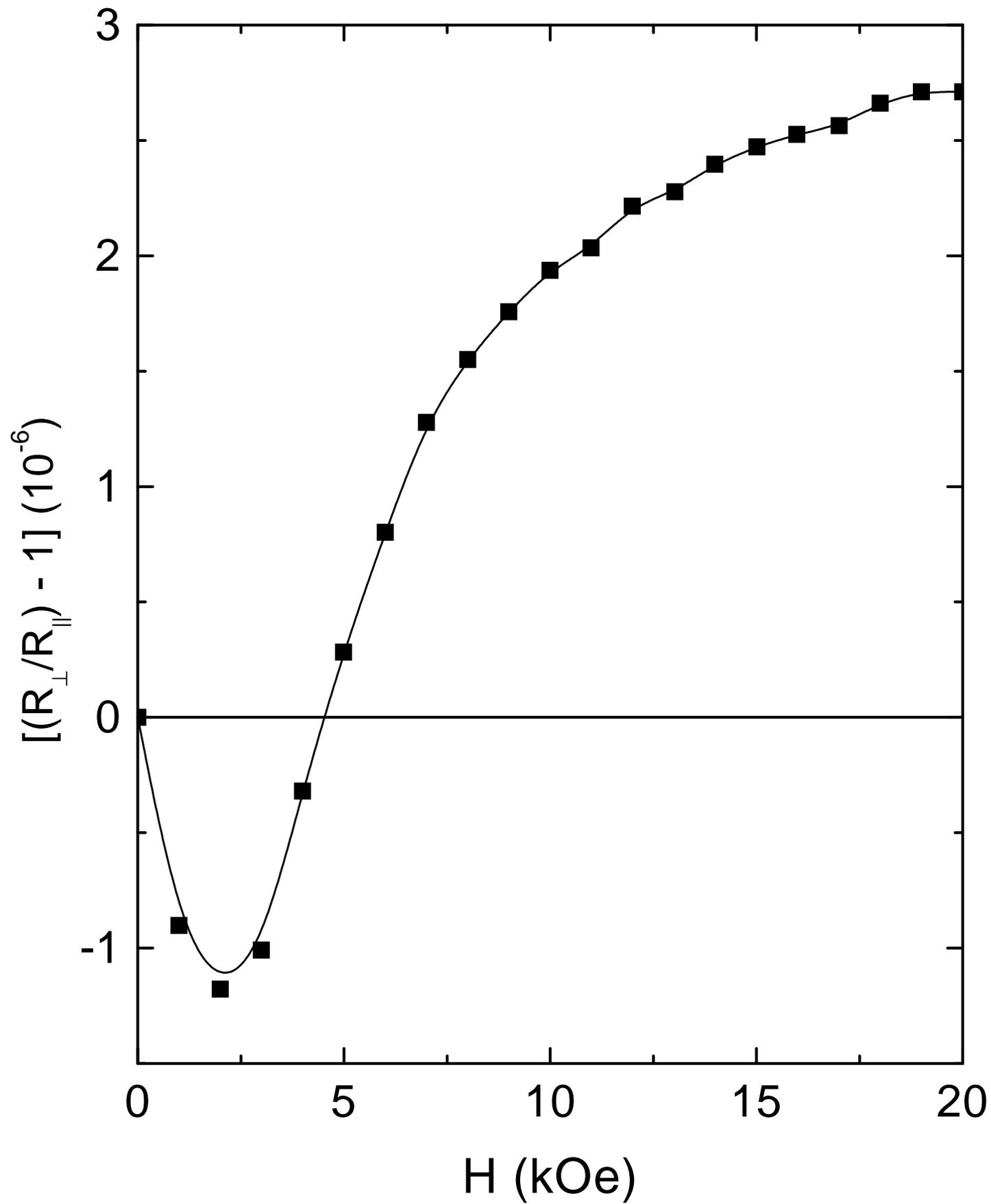
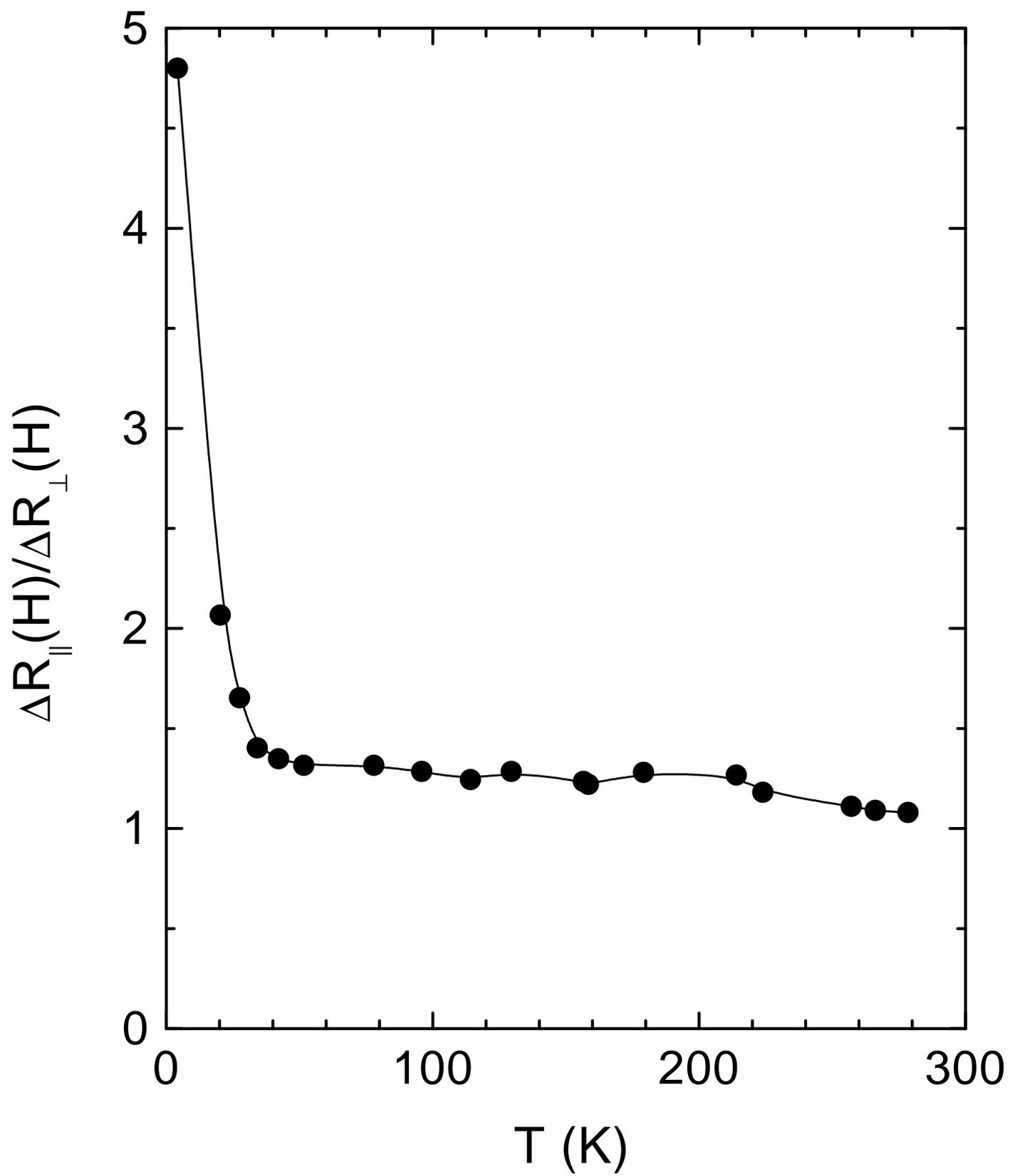
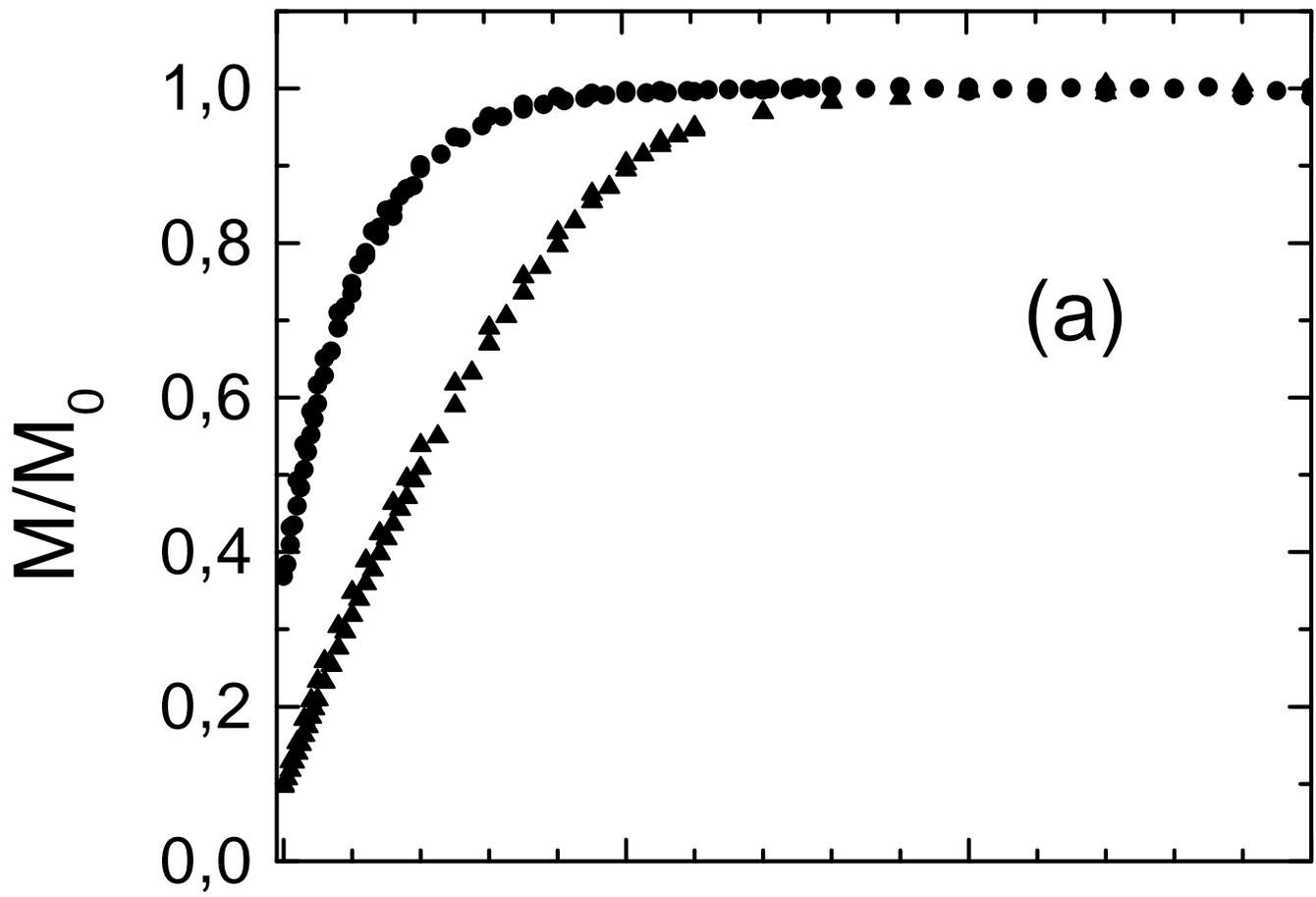
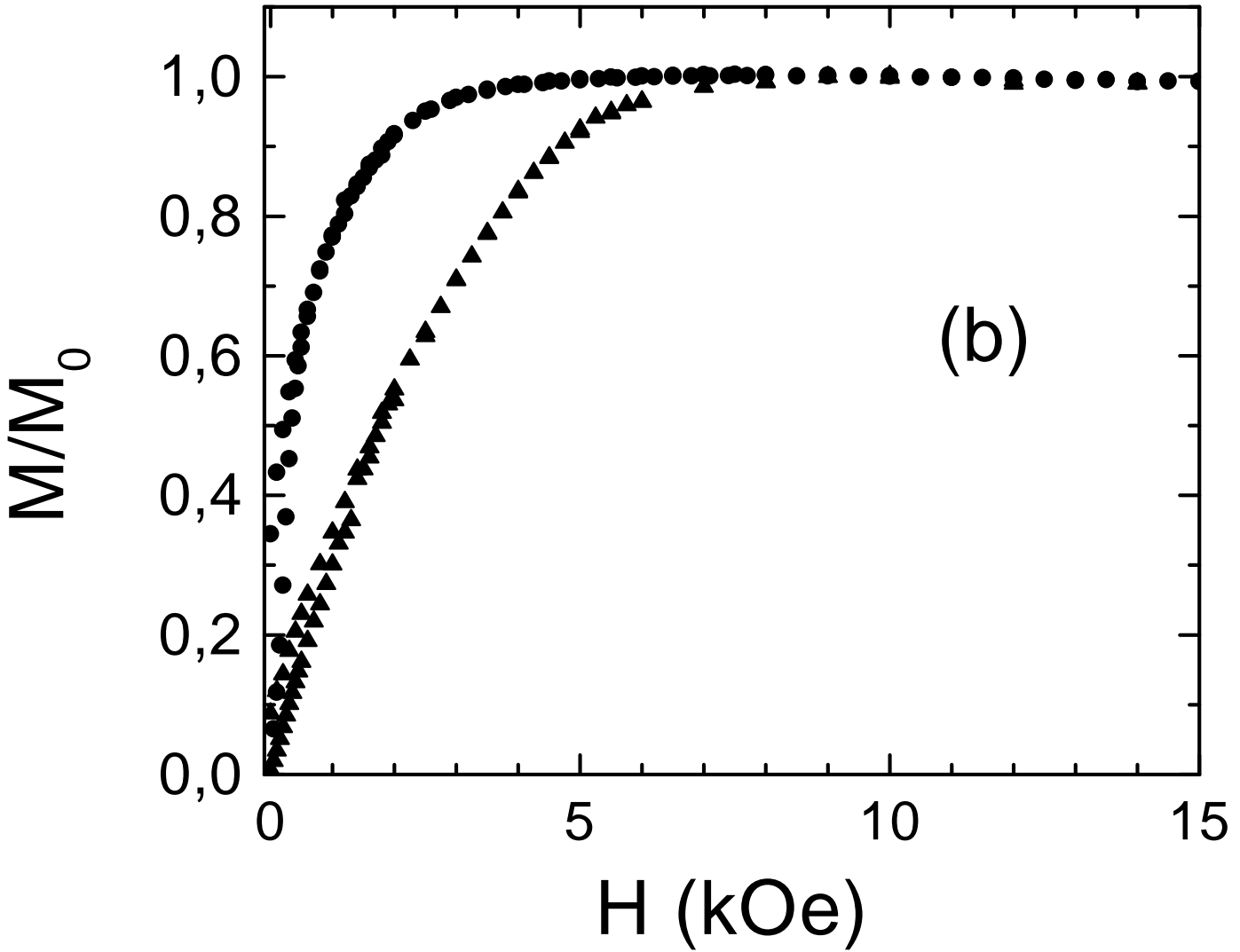
\end{document}